\documentclass[12pt]{article}
\usepackage{amsmath,amssymb,amsthm,amsxtra,overpic,bbm,bm,epsfig,subfigure}
\usepackage{mathrsfs}
\usepackage{graphicx}
\usepackage{epstopdf}
\usepackage{color}
\usepackage{comment}
\usepackage{float}
\usepackage{cite}
\usepackage{slashed,stmaryrd}

\begin{document}

\vspace{0.2cm}
\begin{center}
{\Large\bf Observational Constraints on Secret Neutrino Interactions from Big Bang Nucleosynthesis}
\end{center}
\vspace{0.2cm}

\begin{center}
{\bf Guo-yuan Huang,$^{a, b}$} \footnote{E-mail: huanggy@ihep.ac.cn}
\quad
{\bf Tommy Ohlsson,$^{c, d}$} \footnote{E-mail: tohlsson@kth.se}
\quad
{\bf Shun Zhou$^{a, b, e}$} \footnote{E-mail: zhoush@ihep.ac.cn}
\\
\vspace{0.2cm}
{\small $^a$Institute of High Energy Physics, Chinese Academy of
Sciences, Beijing 100049, China \\
$^b$School of Physical Sciences, University of Chinese Academy of Sciences, Beijing 100049, China \\
$^c$Department of Physics, School of Engineering Sciences, KTH Royal Institute of Technology, \\
AlbaNova University Center, Roslagstullsbacken 21, SE-106 91 Stockholm, Sweden \\
$^d$University of Iceland, Science Institute, Dunhaga 3, IS-107 Reykjavik, Iceland \\
$^e$Center for High Energy Physics, Peking University, Beijing 100871, China}
\end{center}

\vspace{1.5cm}

\begin{abstract}
We investigate possible interactions between neutrinos and massive scalar bosons via $g^{}_{\phi} \overline{\nu} \nu \phi$ (or massive vector bosons via $g^{}_V \overline{\nu} \gamma^\mu \nu V^{}_\mu$) and explore the allowed parameter space of the coupling constant $g^{}_{\phi}$ (or $g^{}_V$) and the scalar (or vector) boson mass $m^{}_\phi$ (or $m^{}_V$) by requiring that these secret neutrino interactions (SNIs) should not spoil the success of Big Bang nucleosynthesis (BBN). Incorporating the SNIs into the evolution of the early Universe in the BBN era, we numerically solve the Boltzmann equations and compare the predictions for the abundances of light elements with observations. It turns out that the constraint on $g^{}_{\phi}$ and $m^{}_\phi$ in the scalar-boson case is rather weak, due to a small number of degrees of freedom. However,  in the vector-boson case, the most stringent bound on the coupling $g^{}_V \lesssim 6\times 10^{-10}$ at $95~\%$ confidence level is obtained for $m^{}_V \simeq 1~{\rm MeV}$, while the bound becomes much weaker $g^{}_V \lesssim 8\times 10^{-6}$ for smaller masses  $m^{}_V \lesssim 10^{-4}~{\rm MeV}$. Moreover, we discuss in some detail how the SNIs affect the cosmological evolution and the abundances of the lightest elements.
\end{abstract}

\newpage
\section{Introduction}

As one of the pillars of the standard model of cosmology, Big Bang nucleosynthesis (BBN) opens a unique window to the early Universe~\cite{Wagoner:1966pv, Boesgaard:1985km, Schramm:1997vs, Olive:1999ij, Iocco:2008va, Cyburt:2015mya} and new physics beyond the standard model of ele\-men\-tary particles~\cite{Sarkar:1995dd, Pospelov:2010hj, Fields:2011zzb}. Based on the standard models of both cosmology and particle physics, the theory of BBN itself essentially contains only one free parameter, i.e., the baryon-to-photon density ratio, which has been determined to be $\eta \equiv n^{}_{\rm b}/n^{}_\gamma = (6.047 \pm 0.074)\times 10^{-10}$ from the precision measurement of the cosmic microwave background (CMB) by the Planck collaboration~\cite{Ade:2015xua}. The observed ratio between the primordial abundance of deuterium and that of hydrogen ${\rm D}/{\rm H}|^{}_{\rm p} = (2.53\pm 0.04)\times 10^{-5}$~\cite{Aver:2015iza, Cooke:2013cba, Bania:2002yj, Sbordone:2010zi}, together with the primordial mass fraction of $^4{\rm He}$, i.e., $Y^{}_{\rm p} \equiv \rho(^4{\rm He})/\rho^{}_{\rm b} = 0.2449 \pm 0.0040$, indicates that $5.7 \times 10^{-10} \lesssim \eta \lesssim 6.7 \times 10^{-10}$ at $95~\%$ confidence level (CL)~\cite{Fields:2014uja, Patrignani:2016xqp}, which is remarkably consistent with the CMB determination of the baryon density.\footnote{The BBN theory can also predict the primordial abundances of ${^3}{\rm He}$ and ${^7}{\rm Li}$. However, the only data on ${^3}{\rm He}$ are available for the solar system and the high-metallicity regions in our Galaxy and it is difficult to infer its primordial abundance~\cite{Fields:2014uja}. On the other hand, the observed relative abundance of lithium is ${^7}{\rm Li}/{\rm H}|^{}_{\rm p} = (1.6\pm 0.3)\times 10^{-10}$, showing a discrepancy in the baryon density between the BBN and CMB estimates~\cite{Fields:2011zzb, Singh:2017jmz}. Since the lithium abundance remains an unresolved issue, it will not be used to draw any constraints in this work.} Therefore, any new physics leading to significant deviations from the standard BBN predictions for the light element abundances will receive restrictive constraints.

In this work, we investigate the observational constraints from BBN on the secret neutrino interaction (SNI) with a massive scalar or vector boson. More explicitly, we consider the SNI only for the left-handed neutrino fields and the relevant Lagrangian can be written as
\begin{equation}
{\cal L}^{}_{\rm SNI} = g^{\alpha \beta}_\phi \overline{\nu^{}_{\alpha{\rm L}}} \nu^{\rm C}_{\beta{\rm L}} \phi + g^{\alpha \beta}_V \overline{\nu^{}_{\alpha{\rm L}}} \gamma^\mu \nu^{}_{\beta{\rm L}} V^{}_\mu + {\rm h.c.} \; ,
\label{eq:L}
\end{equation}
where $\phi$ and $V^{}_\mu$ are the fields for the scalar and vector boson with masses $m^{}_\phi$ and $m^{}_V$, respectively. For simplicity, the coupling constants $g^{\alpha \beta}_\phi$ and $g^{\alpha \beta}_V$ are assumed to be both flavor diagonal and universal for three neutrino flavors, namely, $g^{\alpha \beta}_\phi = g^{}_\phi \delta^{}_{\alpha \beta}$ and $g^{\alpha \beta}_V = g^{}_V \delta^{}_{\alpha \beta}$. The main motivation for such an investigation is two-fold. First, the interaction among neutrinos themselves has never been directly tested in terrestrial experiments, since neutrinos participate only in the neutral-current weak interaction in the standard electroweak theory and there has not been an attempt to collide two neutrino beams. In contrast, the early Universe and the core-collapse supernovae, where the neutrino number density is extremely high and the interaction among themselves is important, serve as ideal places to constrain the SNI~\cite{Dolgov:2002wy, Hannestad:2006zg, Lesgourgues:2006nd, Wong:2011ip, Raffelt:1996wa}. Second, the SNI is expected in many particle-physics models, which have been proposed to generate tiny neutrino masses~\cite{ Chikashige:1980ui, Gelmini:1980re, Choi:1991aa, Acker:1992eh, Georgi:1981pg,Schechter:1981cv} or solve the potential problems associated with dark matter~\cite{Aarssen:2012fx}.

Stringent constraints on the SNI with a massless or massive scalar boson have been derived from observations of the CMB and cosmological large-scale structure formation~\cite{Hannestad:2004qu, Hannestad:2005ex, Bell:2005dr, Basboll:2008fx, Archidiacono:2013dua, Forastieri:2015paa, Cyr-Racine:2013jua, Oldengott:2014qra, Forastieri:2017oma, Lancaster:2017ksf, Oldengott:2017fhy}, the supernova SN1987A~\cite{Kolb:1987qy, Konoplich:1988mj, Farzan:2002wx, Zhou:2011rc, Heurtier:2016otg}, and other experiments~\cite{Ng:2014pca, Ioka:2014kca, Gando:2012pj, Albert:2014fya, Belotsky:2001fb, Laha:2013xua}. These experimental constraints are also applicable with some modifications to the case of a vector boson. To this end, a detailed study of the BBN bounds on the SNI in both the scalar and vector cases is lacking, except for a brief discussion in Ref.~\cite{Ahlgren:2013wba}. For this reason, we now extend the previous work by incorporating the SNI into the cosmological evolution during the BBN era and examining its impact on the light element abundances.

The remaining part of this work is organized as follows. In Sec.~\ref{sec:gf}, we set up the general theoretical framework to study the SNI in the BBN era. A general discussion on how the presence of new particles and interactions affects the standard BBN theory is given. Then, in Sec.~\ref{sec:nr}, we numerically solve the Boltzmann equations for the cosmological evolution and the nucleosynthesis of light elements, where the BBN constraints on the coupling constant and the mass are derived. Finally, in Sec.~\ref{sec:sc}, we summarize our main results and draw our conclusions.

\section{General Framework}
\label{sec:gf}
\subsection{Simple Arguments}
\label{sub:sa}
\begin{figure}[t!]
    \begin{center}
    \includegraphics[width=0.9\textwidth]{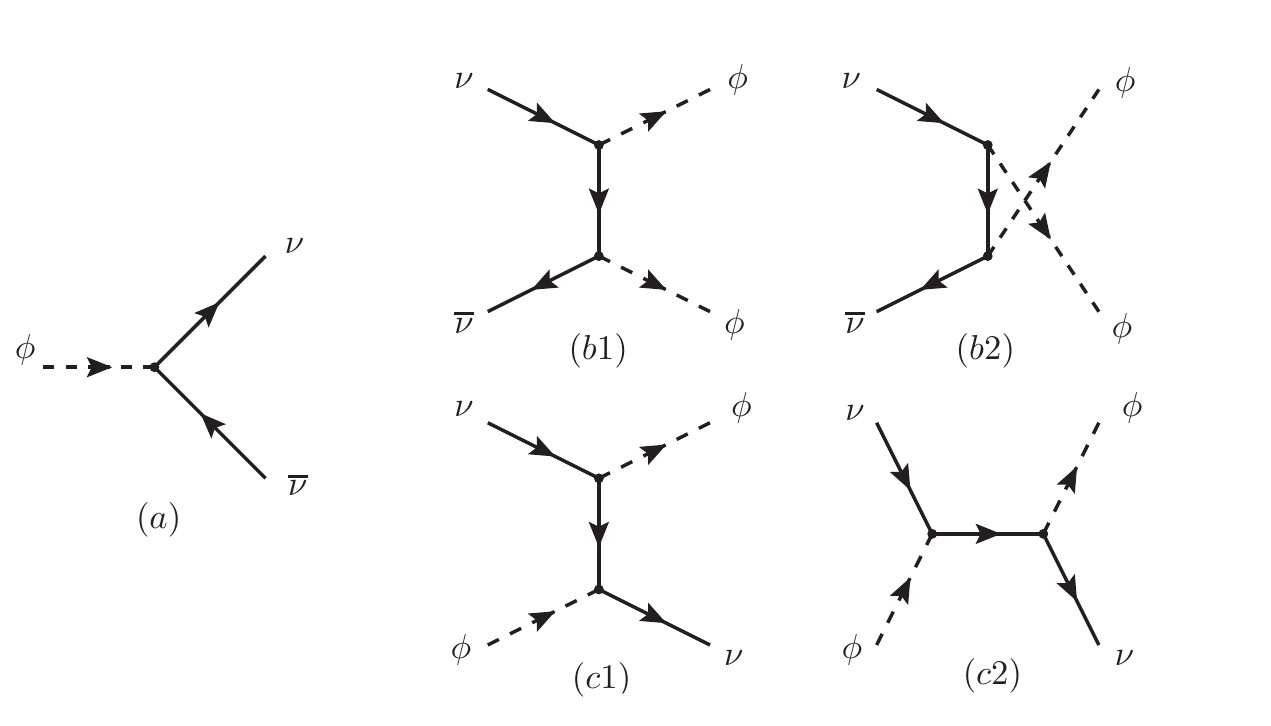}
    \end{center}
\vspace{-0.8cm}
    \caption{The Feynman diagrams for the decay $\phi \to \nu + \overline{\nu}$ (a), the annihilation $\nu + \overline{\nu} \to \phi + \phi$ (b1) and (b2), the elastic scattering $\nu + \phi \to \nu + \phi$ (c1) and (c2) processes in the presence of SNI with a massive scalar boson $\phi$. The corresponding diagrams in the vector case can be obtained by replacing $\phi$ with $V$ accordingly.}
    \label{fig:feynman}
    \end{figure}

As mentioned below Eq.~\eqref{eq:L}, the coupling constant $g^{}_\phi$ or $g^{}_V$ between neutrinos and the new particle $\phi$ or $V$ is assumed to be flavor conserving and universal. The relaxation of such an assumption may lead to slight differences. For instance, if $\phi$ or $V$ is coupled exclusively to $\nu^{}_e$ and copiously produced after the decoupling of $\nu^{}_\mu$ and $\nu^{}_\tau$, the energy density of the decoupled $\nu^{}_\mu$ and $\nu^{}_\tau$ will not be modified, rendering the constraint on the coupling constant relatively weaker. For simplicity, we ignore the flavor dependence of the SNI and treat neutrino flavors equally in the evolution of our Universe.\footnote{If neutrino flavor conversion and non-instantaneous decoupling of neutrinos are taken into account, the effective number $N^{}_{\rm eff}$ of neutrino species (at the temperature far blow $1~{\rm MeV}$) defined via the neutrino-to-photon ratio of energy densities $\rho^{}_\nu/\rho^{}_\gamma = N^{}_{\rm eff}\cdot 7/8 \cdot (4/11)^{4/3}$ will be shifted from $N^{}_{\rm eff} = 3$ to $N^{}_{\rm eff} = 3.045$~\cite{Mangano:2005cc, Birrell:2014uka, Grohs:2015tfy, deSalas:2016ztq}. Such a tiny difference will not be important compared to the radical changes due to the new physics under consideration.} Through the interaction given in Eq.~\eqref{eq:L}, the scalar boson $\phi$ can be generated by the inverse decay $\nu + \overline{\nu} \to \phi$ and the neutrino-antineutrino annihilation $\nu + \overline{\nu} \to \phi + \phi$, for which the Feynman diagrams are shown in Fig.~\ref{fig:feynman}. For the vector boson $V$, we have basically the same production processes. As pointed out in Ref.~\cite{Ahlgren:2013wba}, when $\phi$ or $V$ becomes in thermal equilibrium around the temperature $T \simeq 1~{\rm MeV}$, it contributes to the extra radiation represented in terms of $\Delta N^{}_{\rm eff} \equiv N^{}_{\rm eff} - 3$, where the effective number of neutrino species is defined as $N^{}_{\rm eff} \equiv (\rho^{}_{\rm r} - \rho^{}_\gamma)/(\rho^{\rm std}_\nu/3)$ with $\rho^{}_{\rm r}$ and $\rho^{\rm std}_\nu$ being the energy density of all radiation and the neutrino energy density in terms of the photon temperature $T^{}_{\gamma}$ in the limit of instantaneous decoupling, respectively. With this definition, $N^{}_{\rm eff}$ in the standard case without SNI will be fixed to three during the whole BBN era. In the scalar case with $m^{}_\phi \lesssim 1~{\rm MeV}$, we have only one extra relativistic degree of freedom, corresponding to $\Delta N^{}_{\rm eff} = 1/2\cdot 8/7 \approx 0.57$, whereas in the vector case with $m^{}_V \lesssim 1~{\rm MeV}$, we have three helical states, indicating $\Delta N^{}_{\rm eff} = 3/2\cdot 8/7 \approx 1.71$. It is worthwhile to notice that the temperatures of photons and neutrinos remain equal before the electron-positron annihilation at $T \lesssim 0.511~{\rm MeV}$. If $m^{}_\phi$ or $m^{}_V$ is much larger than $1~{\rm MeV}$, $\phi$ or $V$ is non-relativistic and its number density will be suppressed by the Boltzmann factor $e^{-m^{}_\phi/T}$ or $e^{-m^{}_V/T}$, significantly reducing the contribution to $\Delta N^{}_{\rm eff}$. Then, we require that the upper bound $\Delta N^{}_{\rm eff} < 1$ at $95~\%$ CL~\cite{Mangano:2011ar} should be satisfied for $T \simeq 1~{\rm MeV}$. Obviously, there are essentially no constraints on $g^{}_\phi$ and $m^{}_\phi$, whereas the constraints on $g^{}_V$ and $m^{}_V$ could be restrictive.

In the above discussion, we have only considered the situation in which $V$ can be in thermal equilibrium at $T \simeq 1~{\rm MeV}$, but whether this is the case or not depends on $g^{}_V$ and $m^{}_V$. On the other hand, even if $V$ cannot be brought into thermal equilibrium, it still contributes to the total energy density of our Universe. Therefore, the Boltzmann equations for the distribution functions of both neutrinos and $V$ should be implemented in the latter case to calculate $\Delta N^{}_{\rm eff}$. Before going into details of the Boltzmann equations, we make some comments on the constraints on $g^{}_V$ and $m^{}_V$ by simple dimensional analysis:
\begin{itemize}
\item Around $T \simeq 1~{\rm MeV}$, the Universe is certainly radiation-dominated, so the Hubble expansion rate is given by $H \approx 1.66 \sqrt{g^{}_*} T^2/M^{}_{\rm Pl}$, where $g^{}_* = 10.75$ denotes the effective number of relativistic degrees of freedom and $M^{}_{\rm Pl} \simeq 1.221\times 10^{19}~{\rm GeV}$ is the Planck mass. For $V$ to be thermalized, we have to calculate its production rate $\Gamma^{}_V$ and demand $\Gamma^{}_V \gtrsim H$ at $T \simeq 1~{\rm MeV}$.

\item For a relatively large mass $m^{}_V \lesssim 1~{\rm MeV}$, the inverse decay $\nu + \overline{\nu} \to V$ could be quite efficient, since the decay rate in the rest frame is proportional to both $g^2_V$ and $m^{}_V$. As an order-of-magnitude estimate, we obtain $\Gamma^{\rm D}_V \approx g^2_V m^{}_V/(12\pi) \cdot m^{}_V/(3T)$, where the last factor is the Lorentz factor $\langle E \rangle/m^{}_V \approx 3T/m^{}_V$, arising from a boost to the comoving frame. When the inverse decay dominates the production and brings $V$ into thermal equilibrium, i.e., $\Gamma^{\rm D}_V \gtrsim H$, we arrive at
    \begin{equation}
    g^{}_V \gtrsim 2.2\times 10^{-10} \left(\frac{1~{\rm MeV}}{m^{}_V}\right) \; .
    \label{eq:gV_lowerbound}
    \end{equation}
Consequently, if $V$ is thermalized at $T \simeq 1~{\rm MeV}$ via the inverse decay, the upper bound $\Delta N^{}_{\rm eff} < 1$ can be translated into $g^{}_V \lesssim 2.2\times 10^{-10}~(1~{\rm MeV}/m^{}_V)$. This is well consistent with the result from a more detailed calculation in Ref.~\cite{Ahlgren:2013wba}.

\item For an extremely small mass $m^{}_V \ll 1~{\rm MeV}$, the inverse decay becomes inefficient and the annihilation $\nu + \overline{\nu} \to V + V$ will take over in thermalizing $V$. Since the masses of neutrinos and $V$ are sufficiently small, the only relevant energy scale in question is just $T$. Hence, $\Gamma_V$ can be estimated to $\Gamma^{\rm A}_V \approx g^4_V T$, which should be compared with $H$. Requiring $\Gamma^{\rm A}_V \gtrsim H$ at $T \simeq 1~{\rm MeV}$, we find
    \begin{equation}
    g^{}_V \gtrsim 4.6\times 10^{-6} \; .
    \label{eq:gV_lowerbound2}
    \end{equation}
Similarly, the upper bound $\Delta N^{}_{\rm eff} < 1$ will restrict $g^{}_V$ into the region of $g^{}_V \lesssim 4.6\times 10^{-6}$. An accurate calculation of the total energy density or $\Delta N^{}_{\rm eff}$ in the case of $m^{}_V \ll 1~{\rm MeV}$ has not yet been performed in the literature.
\end{itemize}

The exact calculation of the total energy density for an arbitrary coupling constant $g^{}_\phi$ (or $g^{}_V$) and an arbitrary mass $m^{}_\phi$ (or $m^{}_V$) calls for the implementation of Boltzmann equations. First, both neutrinos and the new boson $\phi$ or $V$ could deviate from the thermal distributions, so the determination of $\Delta N^{}_{\rm eff}$ requires numerical solutions to the true distribution functions. Second, since the BBN takes place for a wide range of temperature (e.g., from $T = 1~{\rm MeV}$ to $T = 0.01~{\rm MeV}$), a naive requirement for $\Delta N^{}_{\rm eff} < 1$ at $T \simeq 1~{\rm MeV}$ oversimplifies the picture of relevant physics.

\subsection{Boltzmann Equations}

In order to fully take into account the decay, annihilation, and scattering processes shown in Fig.~\ref{fig:feynman}, we need to solve a complete set of Boltzmann equations for the distribution functions of neutrinos and the new particle $\phi$ or $V$. The general theoretical framework for the cosmological evolution can be found in a number of excellent books on cosmology (see, e.g., Refs.~\cite{Kolb:1990vq, Dodelson:2003ft, Weinberg:2008zzc}). Therefore, we only outline the strategy for our computations.

First, the Hubble parameter $H(t) \equiv \dot{a}(t)/a(t)$ is governed by the Friedmann equation $H^2 = 8\pi \rho/(3M^2_{\rm Pl})$, where $a(t)$ is the scale factor and $\rho$ is the total energy density. The evolution of the energy density satisfies $\dot{\rho}(t) = -3 H (\rho + P)$, where both $\rho$ and pressure $P$ can be solved for the given particle contents and their distributions. As $\rho = \rho^{}_\gamma + \rho^{}_\nu + \rho^{}_e + \rho^{}_{\rm b} + \rho^{}_{\phi/V}$ is still dominated by radiation in the BBN era, we can safely ignore the contribution $\rho^{}_{\rm b}$ from baryons and count all those from photons ($\gamma$), neutrinos ($\nu$), electrons ($e$), and the new boson $\phi$ (or $V$). This sets up the evolution of the cosmological background.

Second, in the homogeneous and isotropic Universe with the Friedmann--Lema\^{\i}tre--Robertson--Walker metric, the relevant distribution functions $f^{}_i(|{\bf p}^{}_i|, t)$ for $i = \nu$, $\overline{\nu}$, and $\phi$ (or $V$) fulfill the following Boltzmann equations~\cite{Bernstein:1988bw}:
\begin{equation}
\frac{\partial f^{}_i(|{\bf p}^{}_i|, t)}{\partial t} - H |{\bf p}^{}_i| \frac{\partial f^{}_i(|{\bf p}^{}_i|, t)}{\partial |{\bf p}^{}_i|} = C^i_{\rm D}(f^{}_\nu, f^{}_{\phi/V}) + C^i_{\rm A}(f^{}_\nu, f^{}_{\phi/V}) + C^i_{\rm E}(f^{}_\nu, f^{}_{\phi/V}) + C^i_{\rm SM}\; ,
\label{eq:Boltzmann}
\end{equation}
where the quantities $C^i_{\rm D}$, $C^i_{\rm A}$, and $C^i_{\rm E}$ stand for the collision terms of the decay, annihilation, and elastic scattering processes, respectively. In fact, the last term $C^i_{\rm SM}$ in the right-hand side of Eq.~(\ref{eq:Boltzmann}) collectively includes all the relevant scattering processes for neutrinos in the standard model, such as $\nu \nu \leftrightarrow \nu \nu$, $\nu \overline{\nu} \to \nu \overline{\nu}$, $\nu \overline{\nu} \leftrightarrow e^+ e^-$, and $\nu e^- \to \nu e^-$, where the neutrino (antineutrino) flavor indices have been suppressed. Since the neutrino interactions in the standard model have been extensively studied in the literature~\cite{Hannestad:1995rs, Dolgov:1997mb, Esposito:2000hi}, we will not explicitly show them in Eq.~(\ref{eq:Boltzmann}), but have indeed included them in our numerical calculations. For the SNI part, assuming the scalar boson $\phi$, we have
\begin{eqnarray}
C^\phi_{\rm D} &=& \frac{1}{2E^{}_\phi} \int {\rm d}\tilde{p}^{}_\nu {\rm d}\tilde{p}^{}_{\overline{\nu}} \tilde{\delta}^4(p) \left[f^{}_\nu f^{}_{\overline{\nu}} (1 + f^{}_\phi) - f^{}_\phi (1 - f^{}_\nu) (1 - f^{}_{\overline{\nu}})\right] |\overline{\cal M}^{}_{\rm D}|^2 \; , \\
C^\phi_{\rm A} &=& \frac{1}{2E^{}_\phi} \int {\rm d}\tilde{p}^{}_\nu {\rm d}\tilde{p}^{}_{\overline{\nu}} {\rm d}\tilde{p}^\prime_\phi \tilde{\delta}^4(p) \left[f^{}_\nu f^{}_{\overline{\nu}} (1 + f^\prime_\phi) (1 + f^{}_\phi) - f^{}_\phi f^\prime_\phi (1 - f^{}_\nu)(1 - f^{}_{\overline{\nu}})\right] |\overline{\cal M}^{}_{\rm A}|^2 \; , \\
C^\phi_{\rm E} &=& \frac{1}{2E^{}_\phi} \int {\rm d}\tilde{p}^{}_\nu {\rm d}\tilde{p}^\prime_\nu {\rm d}\tilde{p}^\prime_\phi \tilde{\delta}^4(p) \left[f^{}_\nu f^\prime_\phi (1 + f^{}_\phi) (1 - f^\prime_\nu)- f^\prime_\nu f^{}_\phi (1 + f^\prime_\phi)(1 - f^{}_\nu)\right] |\overline{\cal M}^{}_{\rm E}|^2 \; ,
\end{eqnarray}
where ${\rm d}\tilde{p}^{}_i \equiv g^{}_i {\rm d}^3 p^{}_i/[(2\pi)^3 2E^{}_i]$ with $g^{}_i$ being the internal degrees of freedom, $p^{}_i = (E^{}_i, {\bf p}^{}_i)$ denotes the four-momentum, and $\tilde{\delta}^4(p) \equiv (2\pi)^4 \delta^4 (\sum p)$ is the Dirac delta function for four-momentum conservation. The matrix elements squared $|\overline{\cal M}^{}_{\rm D}|^2$, $|\overline{\cal M}^{}_{\rm A}|^2$, and $|\overline{\cal M}^{}_{\rm E}|^2$ are averaged over the initial and final spins. For the annihilation process, one should take care of the symmetric factors, arising from the identical particles in the initial and final states, and the changes of particle numbers in each specific process. Furthermore, the contribution from the elastic scattering between $\overline{\nu}$ and $\phi$ is not explicitly shown in $C^\phi_{\rm E}$, but should be added. In fact, for the Boltzmann equations of neutrinos and antineutrinos, we have also included the standard model processes, which establish a thermal contact between the neutrino sector and the system of photons, electrons, and baryons.

Finally, we come to the nuclear reactions for the generation of light elements. At high temperatures, both neutrons ($n$) and protons ($p$) are in thermal equilibrium, so the neutron-to-proton ratio is given by $n/p = e^{-(m^{}_n - m^{}_p)/T}$. After the weak interaction freezes out and neutrinos decouple from the thermal bath at $T \simeq 1~{\rm MeV}$, we have $n/p \approx 1/6$, which will drop to $1/7$ by the time of nuclear reactions due to beta decays of free neutrons. The first process is the formation of deuterium (${\rm D}$) via $p + n \to {\rm D} + \gamma$, which is at work efficiently after the photo-disintegration rate is suppressed at $T \simeq 0.1~{\rm MeV}$. As neutrons will ultimately be integrated into the most stable light element ${^4}{\rm He}$, one can estimate its mass fraction via $Y^{}_{\rm p} = 2(n/p)/(1 + n/p) \approx 0.25$~\cite{Fields:2014uja}. Although $Y^{}_{\rm p}$ is not quite sensitive to the nuclear reactions, the abundances of ${\rm D}$, ${^3}{\rm He}$, and ${^7}{\rm Li}$ relative to that of ${\rm H}$ are of the order of $10^{-5}$ or even smaller and will be greatly affected by the detailed reactions. In order to numerically calculate $Y^{}_{\rm p}$ and ${\rm D}/{\rm H}|^{}_{\rm p}$, we implement the publicly available code AlterBBN of Ref.~\cite{Arbey:2011nf}, which is actually based on the original Fortran code first presented in Refs.~\cite{Wagoner:1969, Kawano:1992ua} and updated with the latest cross sections of relevant nuclear reactions. Another well-known code PArthENoPE has also been widely used~\cite{Pisanti:2007hk, Consiglio:2017pot}. Note that the neutron lifetime $\tau_n = (880.3 \pm 1.1)~{\rm s}$ will be used in our calculation as an input value~\cite{Patrignani:2016xqp}.

\section{Numerical Results}
\label{sec:nr}
\subsection{Extra Radiation}
    \begin{figure}[t]
    \begin{center}
    \subfigure{%
     \hspace{-0.1cm}
    \includegraphics[width=0.48\textwidth]{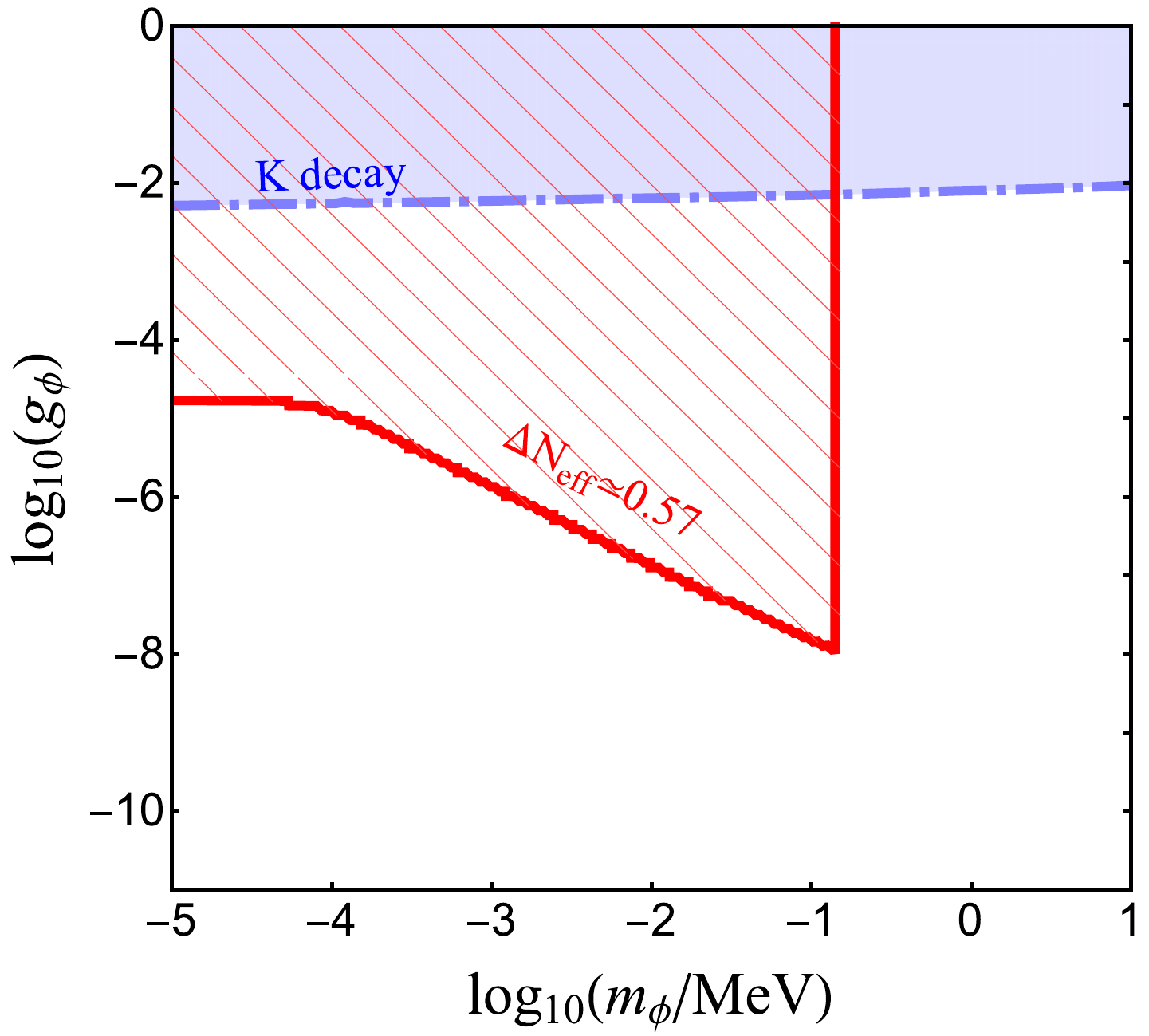}        }%
    \subfigure{%
    \hspace{0.1cm}
    \includegraphics[width=0.48\textwidth]{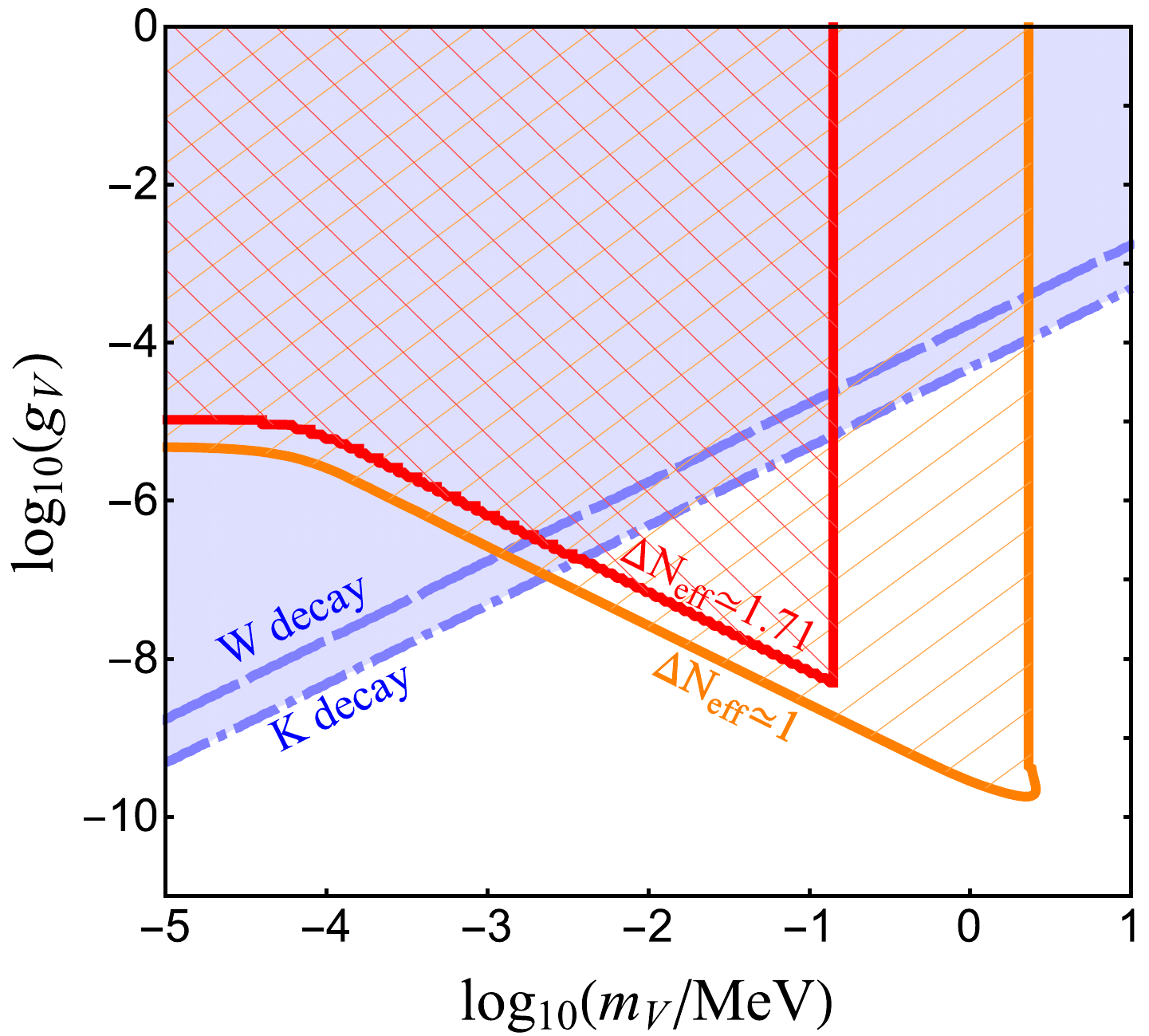}        }
    \end{center}
    \vspace{-0.8cm}
    \caption{The contours of the extra radiation $\Delta N^{}_{\rm eff} \equiv N^{}_{\rm eff} - 3$ at $T^{}_\gamma = 1~{\rm MeV}$ in the plane of $m^{}_\phi$ and $g^{}_\phi$ in the scalar-boson case (left panel) and in the plane of $m^{}_V$ and $g^{}_V$ in the vector-boson case (right panel). The shaded regions with dashed or dotted-dashed lines are excluded by weak decays of kaons and weak gauge bosons, which are reproduced from Ref.~\cite{Laha:2013xua}.}
    \label{fig:phi_contour}
    \end{figure}
As a numerical support for the simple arguments given in Sec.~\ref{sub:sa}, we now attempt to compute the total energy density by solving the Boltzmann equations for the distribution functions of neutrinos and the new particle $\phi$ or $V$ and to determine the extra radiation in terms of $\Delta N^{}_{\rm eff}$ at $T \simeq 1~{\rm MeV}$. For the moment, the nucleosynthesis of light elements is not initiated.

We start by specifying the initial conditions at a high temperature $T = T^{}_\gamma = T^{}_\nu = 10~{\rm MeV}$. For later convenience, $a(t)$ is normalized to $1/T^{}_\gamma$ and two dimensionless parameters $x \equiv m a$ and $q \equiv |{\bf p}| a$ are introduced, where $m$ can be an arbitrary mass scale and is set to $1~{\rm MeV}$ in practice. Photons, neutrinos, and electrons initially follow the distributions in thermal equilibrium with zero chemical potentials, while the initial abundance of $\phi$ or $V$ is assumed to be negligible. With the help of the two dimensionless parameters, we can recast the Boltzmann equations into the following form
\begin{equation}
H x \frac{\partial f^{}_i(q, x)}{\partial x} = C^i_{\rm D}(f^{}_\nu, f^{}_{\phi/V}) + C^i_{\rm A}(f^{}_\nu, f^{}_{\phi/V}) + C^i_{\rm E}(f^{}_\nu, f^{}_{\phi/V}) \; ,
\label{eq:Boltzmann2}
\end{equation}
and obtain $x{\rm d}\rho/{\rm d}x = -3(\rho + P)$, where $\rho$ is in general composed of two parts: the thermal-bath sector $\rho^{}_\gamma$ and $\rho^{}_e$ and the neutrino sector $\rho^{}_\nu$ and $\rho^{}_\phi$ (or $\rho^{}_V$). The former sector can be described by the equilibrium distribution with $T^{}_\gamma$. For the latter sector, the energy density should be calculated from the real distribution functions $f^{}_\nu$ and $f^{}_\phi$ (or $f^{}_V$). There is essentially no difference between neutrinos and antineutrinos, so $\rho^{}_\nu$ actually represents the energy density of both. Then, it is straightforward to derive
\begin{equation}
x\frac{{\rm d}T^{}_\gamma}{{\rm d} x} \frac{{\rm d}(\rho^{}_\gamma + \rho^{}_e)}{{\rm d}T^{}_\gamma} = -3(\rho + P) - x\frac{{\rm d} \rho^{}_\nu}{{\rm d}x}  - x \frac{{\rm d}\rho^{}_{\phi/V}}{{\rm d}x} \; .
\label{eq:Boltzmann3}
\end{equation}
Solving Eqs.~\eqref{eq:Boltzmann2} and \eqref{eq:Boltzmann3} with $H = 8\pi \rho/(3M^2_{\rm Pl})$, we obtain the total energy density of radiation for any given values of $g^{}_\phi$ (or $g^{}_V$) and $m^{}_\phi$ (or $m^{}_V$) and can extract the extra radiation $\Delta N^{}_{\rm eff}$.

In Fig.~\ref{fig:phi_contour}, the numerical results are presented, where the contours of $\Delta N^{}_{\rm eff}$ have been plotted in the plane of $m^{}_\phi$ and $g^{}_\phi$ in the scalar-boson case (left panel) and in the plane of $m^{}_V$ and $g^{}_V$ in the vector-boson case (right panel). Compared with the previous study in Ref.~\cite{Ahlgren:2013wba}, the parameter space has now been extended to the region of much smaller masses, for which the annihilation processes become dominant in the production of $\phi$ or $V$. Some interesting features can be observed from Fig.~\ref{fig:phi_contour}. First, in both plots, one can see that the contours turn out to be flat at the small-mass end, e.g., $m^{}_\phi$ or $m^{}_V \approx 10^{-5}~{\rm MeV}$, indicating that the results of $\Delta N^{}_{\rm eff}$ can be simply extrapolated to the cases of even smaller masses. Second, at the high-mass end, even if $\phi$ or $V$ can be in thermal equilibrium, the Boltzmann factor leads to a suppression of $\Delta N^{}_{\rm eff}$. This is why $\Delta N^{}_{\rm eff}$ decreases quickly when the mass increases. Third, in between low and high masses, the product of $g^{}_\phi m^{}_\phi$ or $g^{}_V m^{}_V$ is nearly constant for a given $\Delta N^{}_{\rm eff}$, which can be understood by the simple estimate in Eq.~\eqref{eq:gV_lowerbound}. However, the maximum of $\Delta N^{}_{\rm eff}$ in the scalar case is $0.57$, when $\phi$ is relativistic and the SNI can bring it into thermal equilibrium with neutrinos. As a consequence, the BBN bound $\Delta N^{}_{\rm eff} \lesssim 1$ from Ref.~\cite{Mangano:2011ar} does not have any constraining power on $g^{}_\phi$ and $m^{}_\phi$.

\subsection{Light Element Abundances}
	 \begin{figure}[t!]
    \begin{center}
    \subfigure{%
    \hspace{-0.5cm}
    \includegraphics[width=0.46\textwidth]{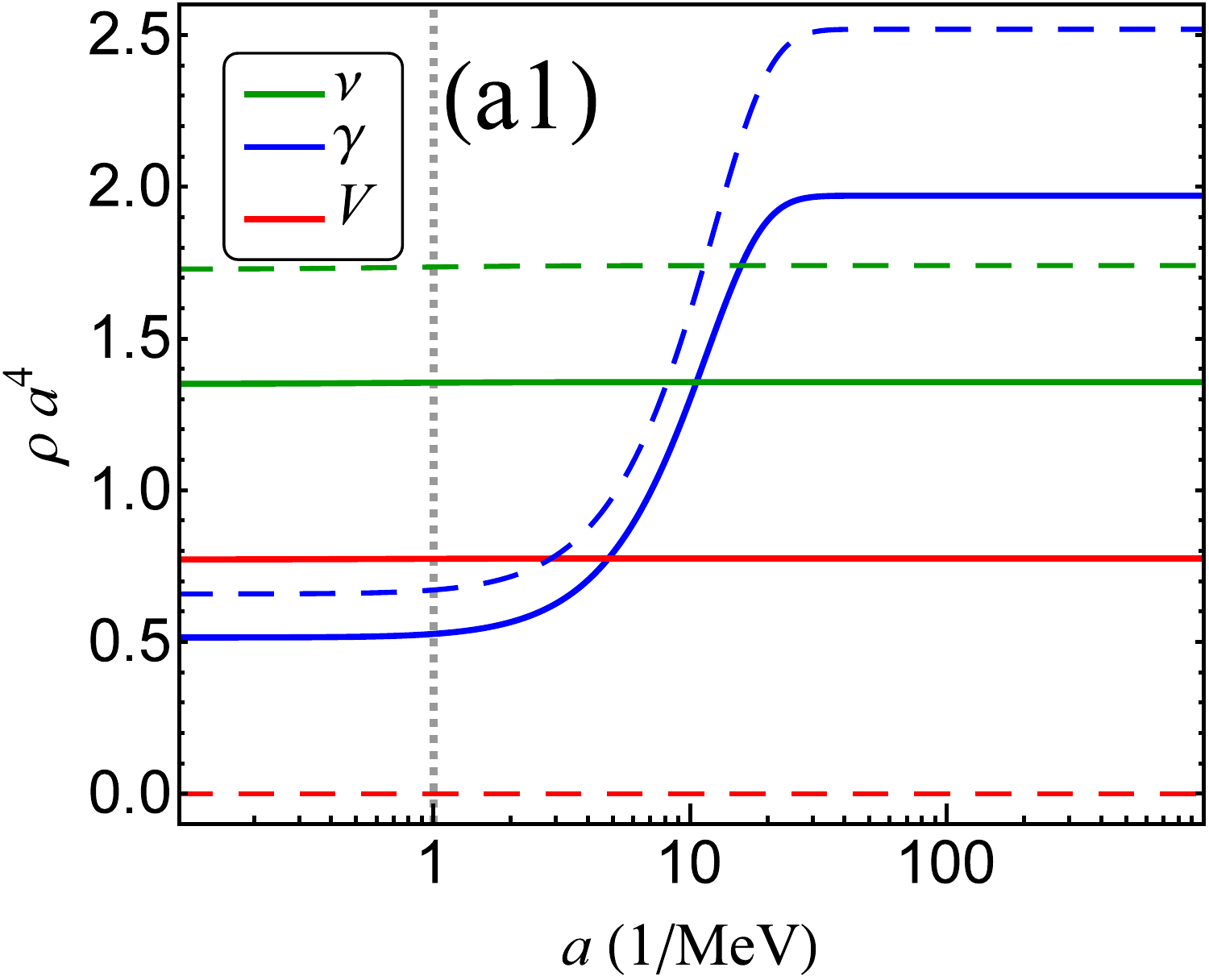}        }
    \subfigure{%
    \hspace{0.5cm}
    \includegraphics[width=0.43\textwidth]{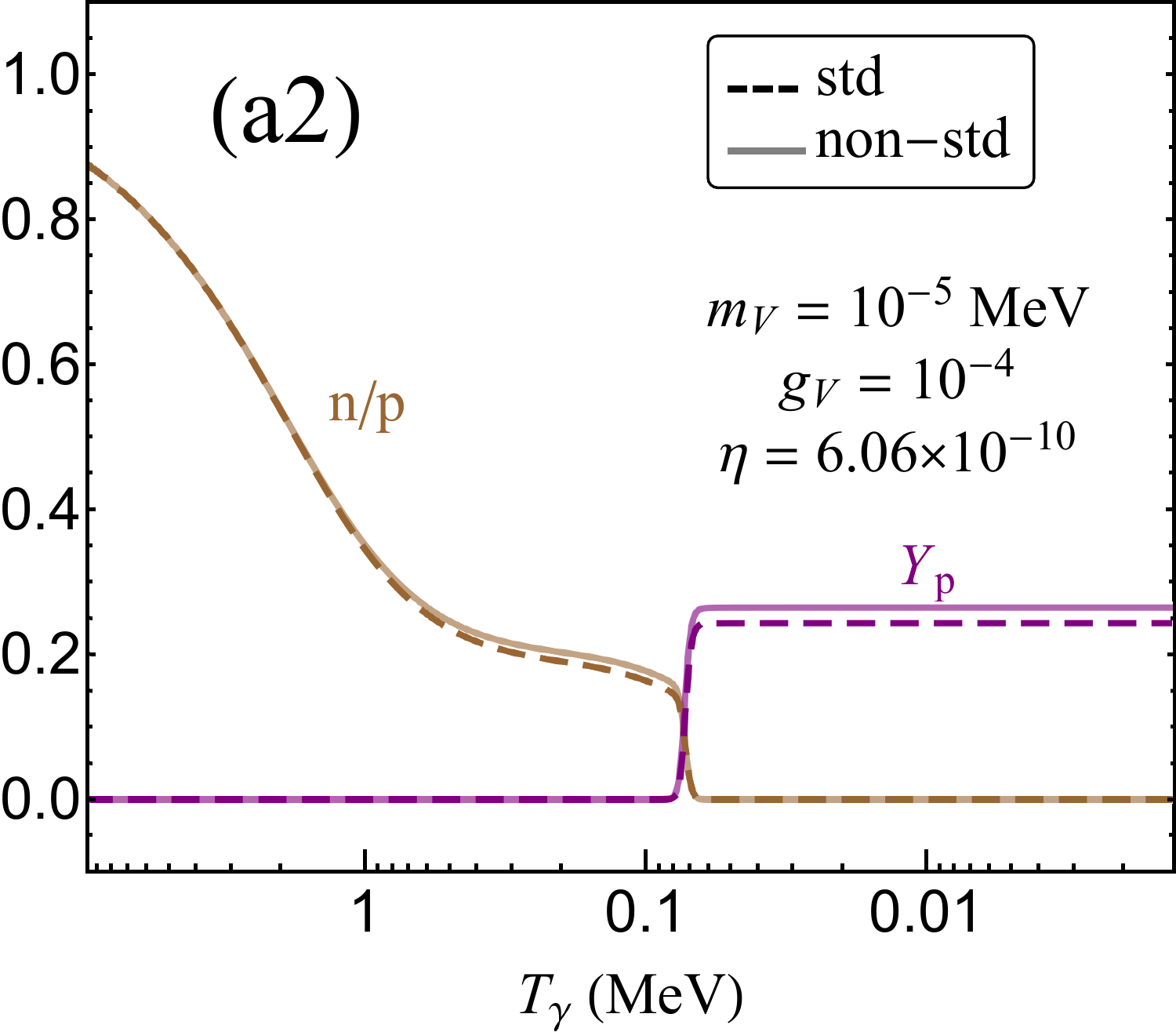}
         }
    \subfigure{%
    \hspace{-0.5cm}
    \includegraphics[width=0.46\textwidth]{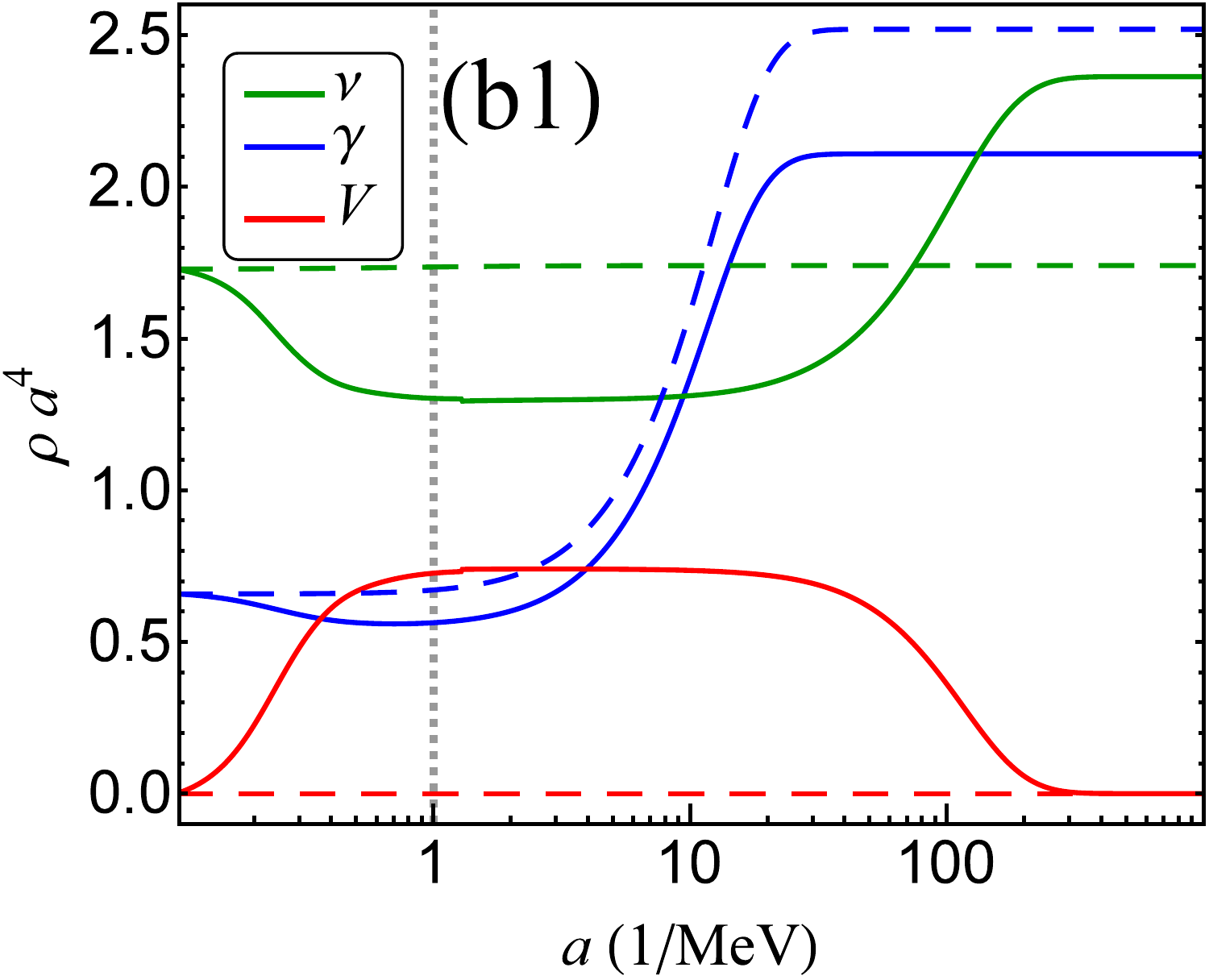}        }
    \subfigure{%
    \hspace{0.5cm}
    \includegraphics[width=0.43\textwidth]{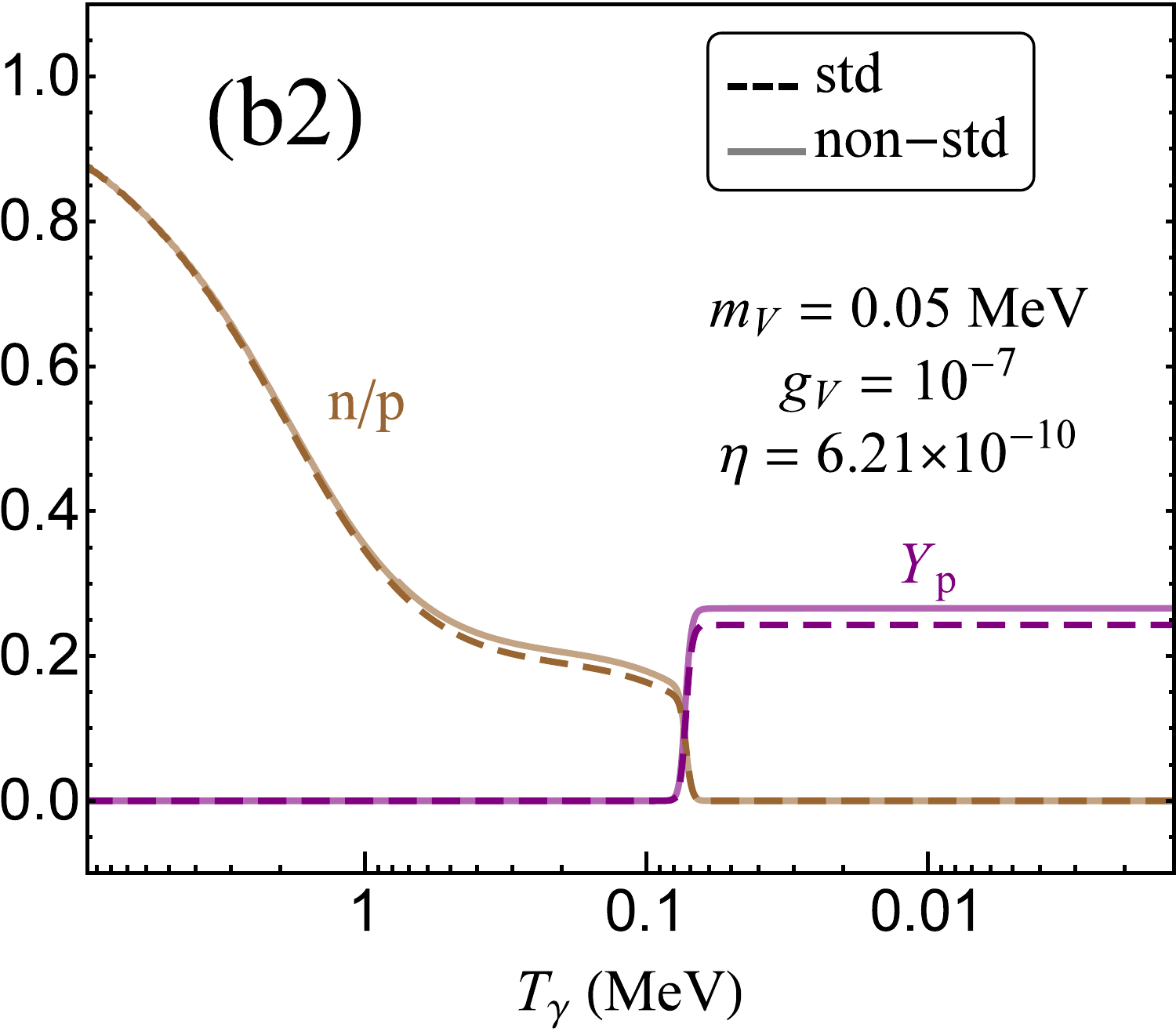}
         }
    \end{center}
    \vspace{-0.8cm}
    \caption{The cosmological evolution of the comoving energy densities for neutrinos $\nu$ (green curves), photons $\gamma$ (blue curves), and vector bosons $V$ (red curves) is presented in the left panel and that of the neutron-to-proton ratio $n/p$ (brown curves) and the helium mass fraction $Y^{}_{\rm p}$ (purple curves) in the right panel. The plots (a1) and (a2) are given for $g^{}_V = 10^{-4}$ and $m^{}_V = 10^{-5}~{\rm MeV}$ in Case~I, while (b1) and (b2) for $g^{}_V = 10^{-7}$ and $m^{}_V = 0.05~{\rm MeV}$ in Case~II. Note that the evolution of the comoving energy densities is plotted with respect to the scale factor $a(t)$ that has been normalized to $1/T^{}_{\gamma}$ at very high temperatures. The dotted vertical lines at $a = 1/{\rm MeV}$ or $T^{}_\gamma = 1~{\rm MeV}$ in the plots (a1) and (b1) represent basically the epoch of neutrino decoupling. The dashed curves are for the standard theory, while the solid curves for the non-standard one. The baryon-to-photon ratio $\eta$ takes on the value, which minimizes the $\chi^2$ function for each case.}
    \label{fig:evolution}
    \end{figure}
     \begin{figure}[t!]
    \begin{center}
    \subfigure{%
    \hspace{-0.5cm}
    \includegraphics[width=0.46\textwidth]{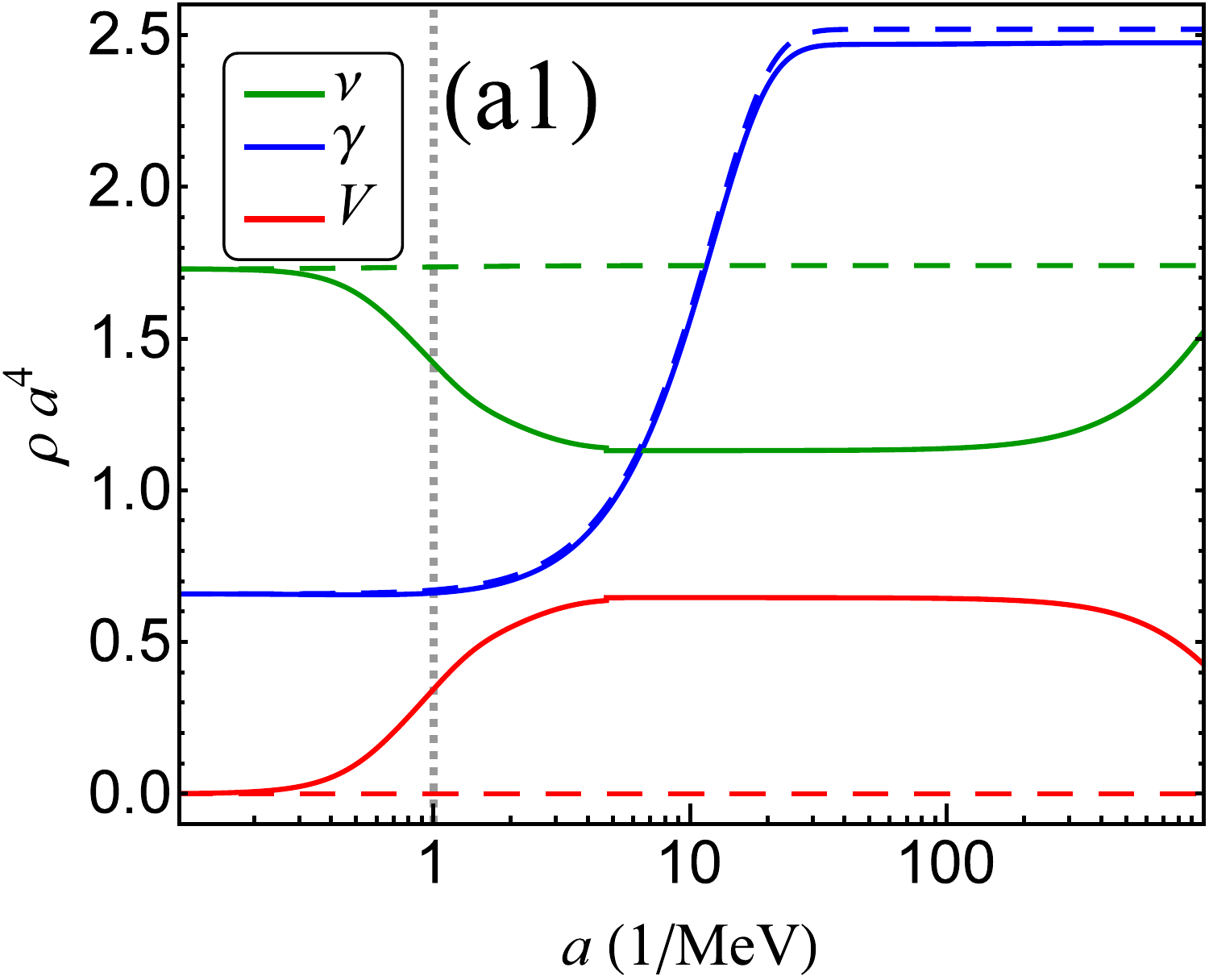}        }
    \subfigure{%
    \hspace{0.5cm}
    \includegraphics[width=0.43\textwidth]{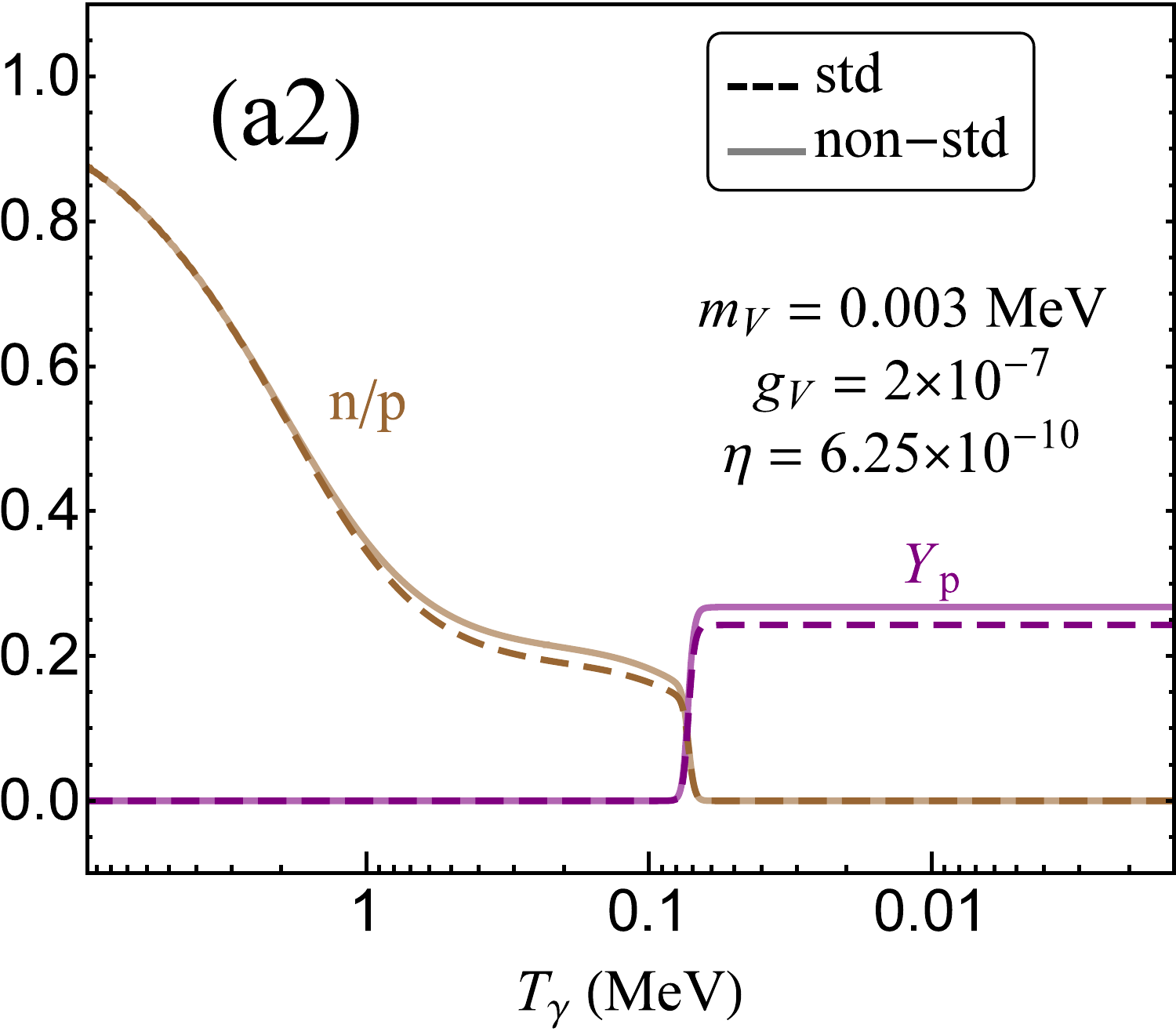}
         }
    \subfigure{%
    \hspace{-0.5cm}
    \includegraphics[width=0.46\textwidth]{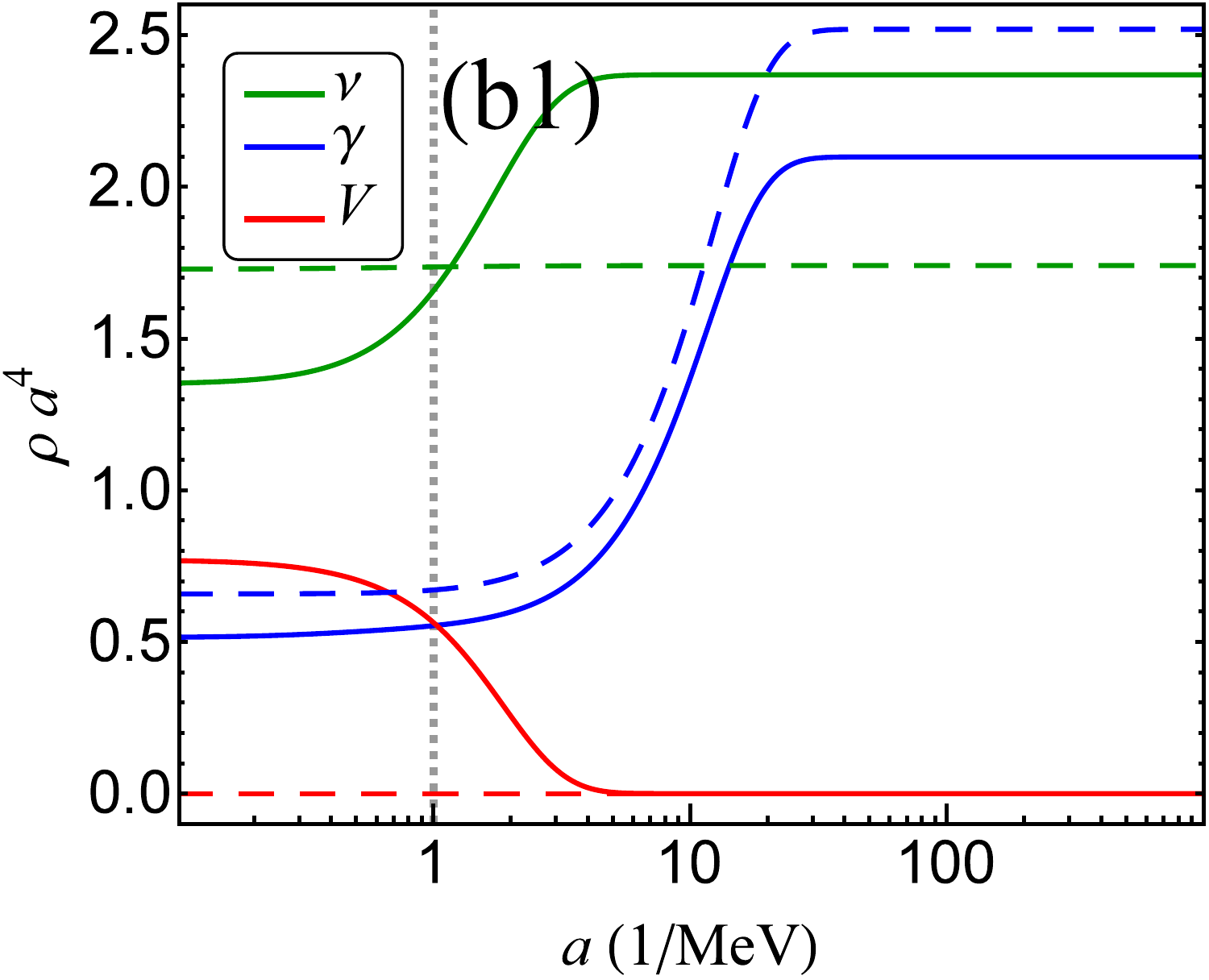}        }
    \subfigure{%
    \hspace{0.5cm}
    \includegraphics[width=0.43\textwidth]{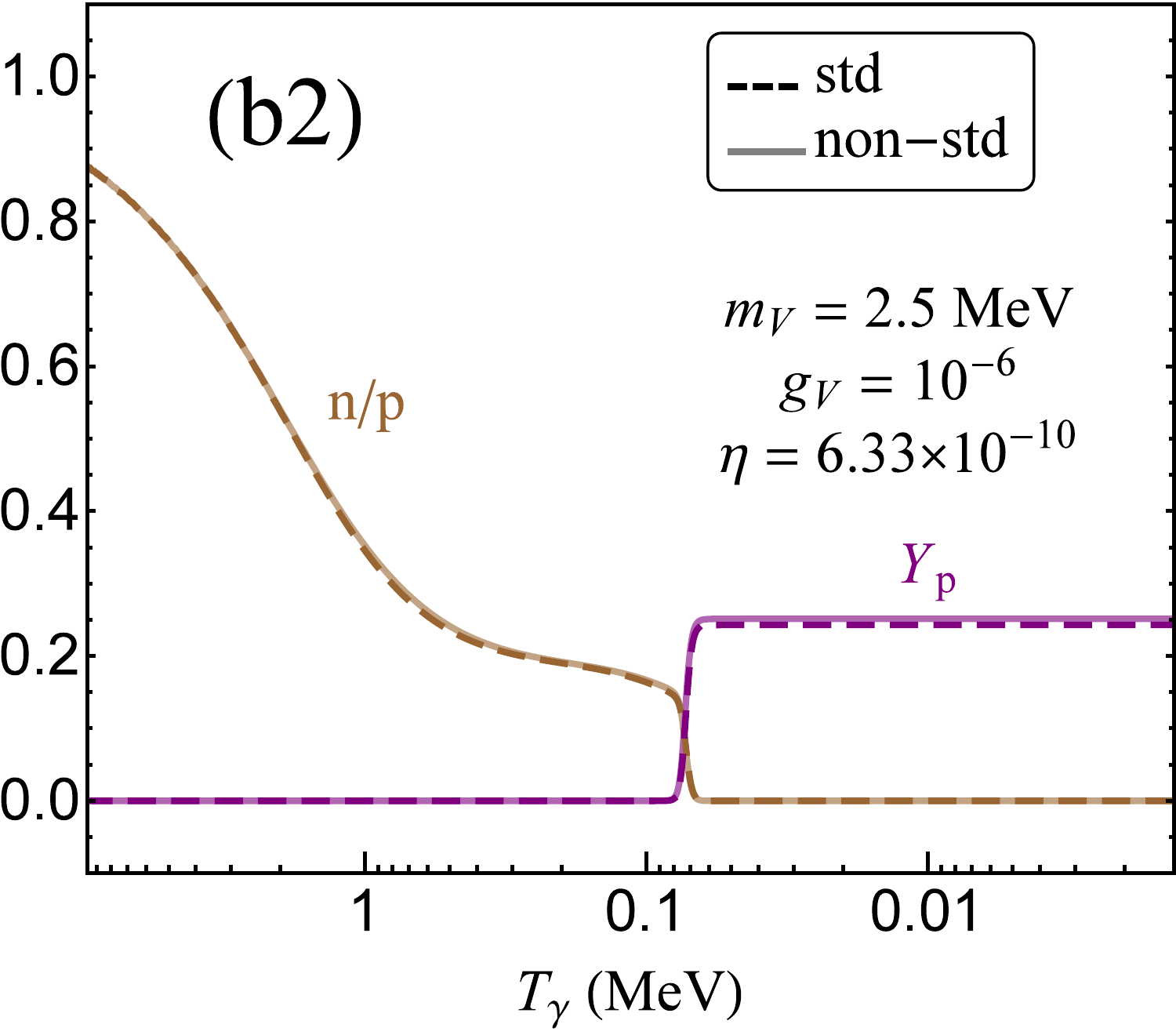}
         }
    \end{center}
    \vspace{-0.8cm}
    \caption{The cosmological evolution of the comoving energy densities for neutrinos $\nu$ (green curves), photons $\gamma$ (blue curves), and vector bosons $V$ (red curves) is presented in the left panel and that of the neutron-to-proton ratio $n/p$ (brown curves) and the helium mass fraction $Y^{}_{\rm p}$ (purple curves) in the right panel. The plots (a1) and (a2) are given for $g^{}_V = 2\times 10^{-7}$ and $m^{}_V = 0.003~{\rm MeV}$ in Case~III, while (b1) and (b2) for $g^{}_V = 10^{-6}$ and $m^{}_V = 2.5~{\rm MeV}$ in Case~IV. The notations and other input parameters are the same as for Fig.~\ref{fig:evolution}.}
    \label{fig:evolution2}
    \end{figure}
After having numerically computed the extra radiation, we proceed with the real evolution of light element abundances by combining the Boltzmann equations with the AlterBBN code~\cite{Arbey:2011nf}. Solving the Boltzmann equations numerically, we can obtain the necessary information on the evolution of the cosmic background. Such an information will be input to AlterBBN as an alternative model of cosmology, and thus, the light element abundances can be calculated at any instant of time or temperature.

As have already been observed in previous sections, $\phi$ cannot contribute significantly to an extra radiation during the BBN era. In order to illustrate the main effects of the SNI in the BBN, we will concentrate on $V$ and try to capture the most important points by analyzing four different cases of couplings and masses.

{\bf Case I:} $g^{}_V = 10^{-4}$ and $m^{}_V = 10^{-5}~{\rm MeV}$. In Fig.~\ref{fig:evolution} (a1), we show the evolution of the comoving energy densities of $\gamma$'s, $\nu$'s, and $V$'s. In Fig.~\ref{fig:evolution} (a2), the neutron-to-proton ratio $n/p$ and the mass fraction of helium $Y^{}_{\rm p}$ are given. The standard theory (dashed curves) is rather simple, which means that the comoving energy density of $\gamma$'s increases due to the heating from electron-positron annihilation, while that of $\nu$'s remains nearly unchanged, since $\nu$'s are almost decoupled from $\gamma$'s after $1~{\rm MeV}$. In contrast, in this non-standard case, the coupling constant $g^{}_V = 10^{-4}$ is so strong that the $V$'s have been tightly coupled with $\nu$'s even before $T^{}_\gamma = 10~{\rm MeV}$ or $a = 0.1/{\rm MeV}$. The non-standard (solid blue curve) comoving energy density of $\gamma$'s is smaller than the standard one (dashed blue curve), since the entropy has been transferred into $V$'s that are abundantly produced. The same behavior of the comoving energy density of $\nu$'s (green curves) is also found. However, one should notice that the physical processes actually depend on $T^{}_\gamma$ or $\rho^{}_\gamma$ instead of $a(t)$. One can also observe that the energy density ratio of the extra degree of freedom to $\gamma$'s by far exceeds that of the standard case. Namely, the extra radiation in terms of the effective number of neutrino species is $N^{}_{\rm eff} = 4.71$. Consequently, this would accelerate the expansion of the Universe, leading to an earlier decoupling of the weak interaction that interchanges neutrons and protons and a larger value of $n/p$ would then remain for ${}^4{\rm He}$ synthesis. This can been observed in Fig.~\ref{fig:evolution} (a2), where the asymptotic value of $Y^{}_{\rm p}$ is $8.5~\%$ larger than the standard value. For comparison, the $1\sigma$ error of the observed $Y^{}_{\rm p}$ is only $1.6~\%$. Therefore, this case should give a restrictive constraint.

{\bf Case II:} $g^{}_V = 10^{-7}$ and $m^{}_V = 0.05~{\rm MeV}$. In Fig.~\ref{fig:evolution} (b1) and (b2), the numerical results for this case are given. Since $g^{}_V = 10^{-7}$ is now much smaller compared to that in Case~I, the comoving energy densities of $\gamma$'s (blue curves) and $\nu$'s (green curves) coincide with those of the standard values at $T^{}_\gamma = 10~{\rm MeV}$. Shortly later on, one can observe sizable deviations due to the production of $V$'s. Around $a \approx 50/{\rm MeV}$ or $T^{}_\gamma \approx 0.02~{\rm MeV}$ when the $V$'s remain in thermal equilibrium with $\nu$'s, the comoving energy density $\rho^{}_V$ (red curves) is suppressed by the Boltzmann factor, since $m^{}_V = 0.05~{\rm MeV} > T^{}_\gamma \approx 0.02~{\rm MeV}$. The reduction of the number density of $V$'s is taking place via both the decay $V \to \nu + \overline{\nu}$ and the annihilation $V + V \to \nu + \overline{\nu}$ processes. This is the reason why the comoving energy density of $\nu$'s takes over that of $\gamma$'s after these processes are almost completed. Since the $V$'s are thermally populated already by the time of neutrino decoupling at $T^{}_\gamma = 1~{\rm MeV}$, $\Delta N^{}_{\rm eff}$ is the same as in Case~I and its impact on $n/p$ and $Y^{}_{\rm p}$ is also quite similar.
\begin{figure}[t!]
    \begin{center}
    \includegraphics[width=0.48\textwidth]{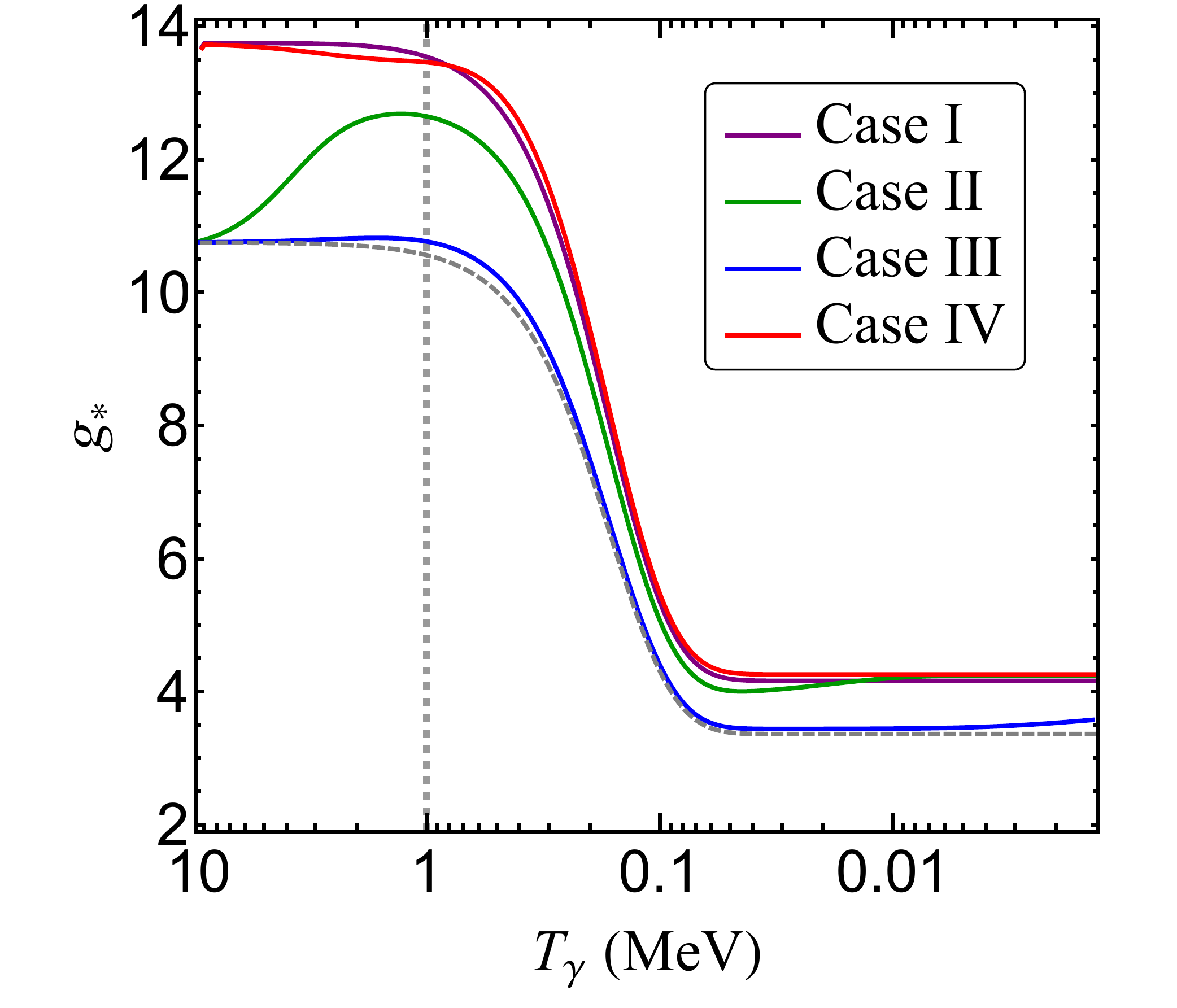}
    \includegraphics[width=0.48\textwidth]{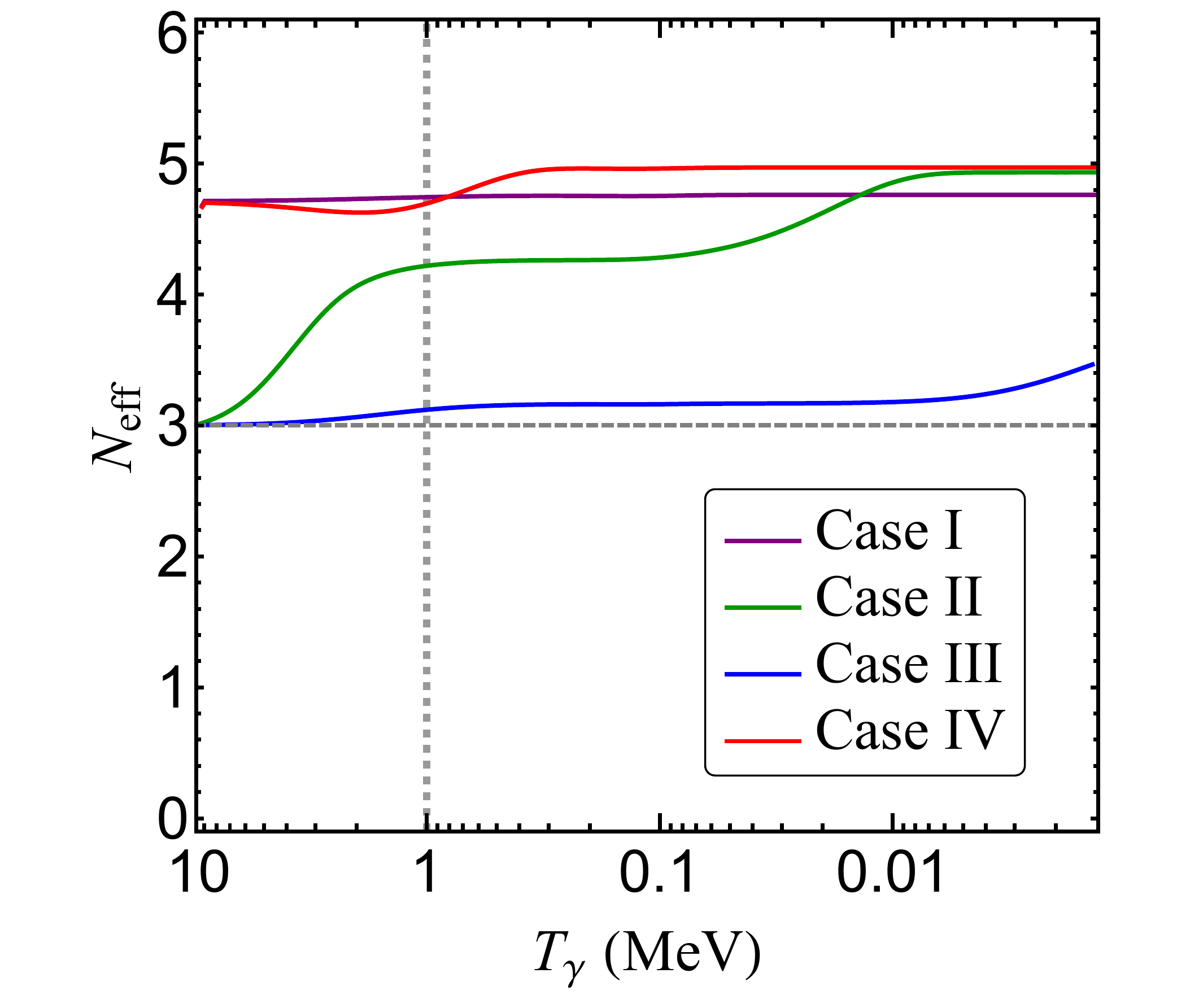}
    \end{center}
	\vspace{-0.8cm}
    \caption{The evolutions of $g^{}_*$ (left panel) and $N^{}_{\rm eff}$ (right panel) with respect to $T_{\gamma}$ for different representative parameters: {\bf Case I}: $g^{}_V = 10^{-4}$ and $m^{}_V = 10^{-5}~{\rm MeV}$ (purple curve), {\bf Case II}: $g^{}_V = 10^{-7}$ and $m^{}_V = 0.05~{\rm MeV}$ (green curve), {\bf Case III}: $g^{}_V = 2\times 10^{-7}$ and $m^{}_V = 0.003~{\rm MeV}$ (blue curve), and {\bf Case IV}: $g^{}_V = 10^{-6}$ and $m^{}_V = 2.5~{\rm MeV}$ (red curve). The dashed curve (horizontal dashed line) in the left (right) panel denotes the result in the standard case with instantaneous decoupling, while the vertical dotted line at $T^{}_{\gamma} = 1~{\rm MeV}$ represents simply the epoch of neutrino decoupling.}
    \label{fig:Neff}
    \end{figure}

{\bf Case III:} $g^{}_V = 2\times 10^{-7}$ and $m^{}_V = 0.003~{\rm MeV}$. In Fig.~\ref{fig:evolution2} (a1) and (a2), the numerical results for this case are displayed. From Fig.~\ref{fig:evolution2} (a1), one can observe that the $V$'s (red curves) are mostly generated after the neutrino decoupling at $T^{}_\gamma \approx 1~{\rm MeV}$, rendering the energy density of $\gamma$'s (blue curves) to be nearly unchanged. In this case, the extra radiation is found to be $\Delta N^{}_{\rm eff} \approx 0.5$. However, $T^{}_\nu$ is severely reduced by thermalizing the $V$'s. This picture cannot be effectively described by $\Delta N^{}_{\rm eff}$. As indicated in Fig.~\ref{fig:evolution2} (a2), the increase of $n/p$ and $Y^{}_{\rm p}$ is more significant than those in the two previous cases, although $\Delta N^{}_{\rm eff}$ is smaller. The true reason is that the lower $T^{}_\nu$ would cause an earlier decoupling of the weak interaction, implying a larger value of $n/p$. In this case, we find that the final $Y^{}_{\rm p}$ deviates from the standard value by $10~\%$. This actually results in the peak of $\Delta \chi^2$ on the edge of the excluded region in Fig.~\ref{fig:bbn_constraints}, which will be discussed in the next subsection.

{\bf Case IV:} $g^{}_V = 10^{-6}$ and $m^{}_V = 2.5~{\rm MeV}$. In Fig.~\ref{fig:evolution2} (b1) and (b2), the numerical results for this case are shown. Due to the large mass, the $V$'s are mainly produced via the inverse decay $\nu + \overline{\nu} \to V$ and have been tightly coupled to $\nu$'s before $T^{}_\gamma = 10~{\rm MeV}$. In this case, the extra radiation is about $\Delta N^{}_{\rm eff} \approx 1.5$. However, the light element abundances seem to be unaffected, as indicated in Fig.~\ref{fig:evolution2} (b2). This can be interpreted as a cancellation between the effects of a higher $T^{}_\nu$ and a larger expansion rate. One may extend the discussion on this case to future works, since this might provide a possible solution to the ${^7}{\rm Li}$ abundance problem~\cite{Fields:2011zzb}, which is, however, beyond the scope of the present work.

In Fig.~\ref{fig:Neff}, the evolutions of $g^{}_*$ and $N^{}_{\rm eff}$ as functions of $T^{}_\gamma$ have been displayed for different parameters in the vector-boson case. As we have mentioned before, $N^{}_{\rm eff}$ has directly been calculated from the true distribution functions of $\nu$'s and $V$'s, which themselves are found by solving the relevant Boltzmann equations. Since $N^{}_{\rm eff}$ itself does evolve with respect to temperature, it is not correct to just draw the constraints on $g^{}_V$ and $m^{}_V$ by simply putting an upper bound on $N^{}_{\rm eff}$ at any instant of temperature or time. Although the $V$'s contribute to $N^{}_{\rm eff}$ by at most 1.71 when they are relativistic and in thermal equilibrium, $N^{}_{\rm eff} \approx 5$ can be reached in Cases~II and IV, as shown in the right panel of Fig.~\ref{fig:Neff}. The main reason is that the $V$'s decay into $\nu$'s, when $T^{}_\gamma$ drops below $m^{}_V$, and transfer their energies into $\nu$'s. This feature can also be observed from Figs.~\ref{fig:evolution} (b1) and \ref{fig:evolution2} (b1). Since the $V$'s are non-relativistic at this stage, their energy densities will be diluted less significantly than those of relativistic particles, such as $\nu$'s. Therefore, the total energy density in terms of $N^{}_{\rm eff}$ could go slightly beyond the value $4.71$. For comparison, in the left panel of Fig.~\ref{fig:Neff}, we also present $g^{}_*(T^{}_\gamma)$, which is defined through the radiation energy density $\rho^{}_{\rm r}(T^{}_\gamma) = \pi^2 g^{}_*(T^{}_\gamma) T^4_\gamma/30$.

From the above discussion of the four different cases, it is now clear how the light element abundances are affected by the SNI with a massive scalar or vector boson. The impact of a scalar boson is moderate, since it has a smaller number of degrees of freedom compared to a vector boson. In general, the introduction of the SNI will produce additional radiation to accelerate the expansion of the Universe. In addition, the thermal contact between the new particle $\phi$ or $V$ and neutrinos will change $T^{}_\nu$ after neutrino decoupling. Both effects will be important for the evolution of the light element abundances. It should be noticed that the abundance of deuterium is also modified when the SNI is present, but the deviations from the standard theory for Cases~I--IV are within $1~\%$ when the $\chi^2$ function is minimized with respect to $\eta$. Thus, the evolution of deuterium is not included. Details on the minimization of $\chi^2$ can be found in the next subsection, where more discussions on the evolution of light element abundances will be given.

\subsection{Final Constraints}
	 \begin{figure}[t]
  	     \vspace{-1cm}
    \begin{center}
    \subfigure{%
    \hspace{-1cm}
    \includegraphics[width=0.65\textwidth]{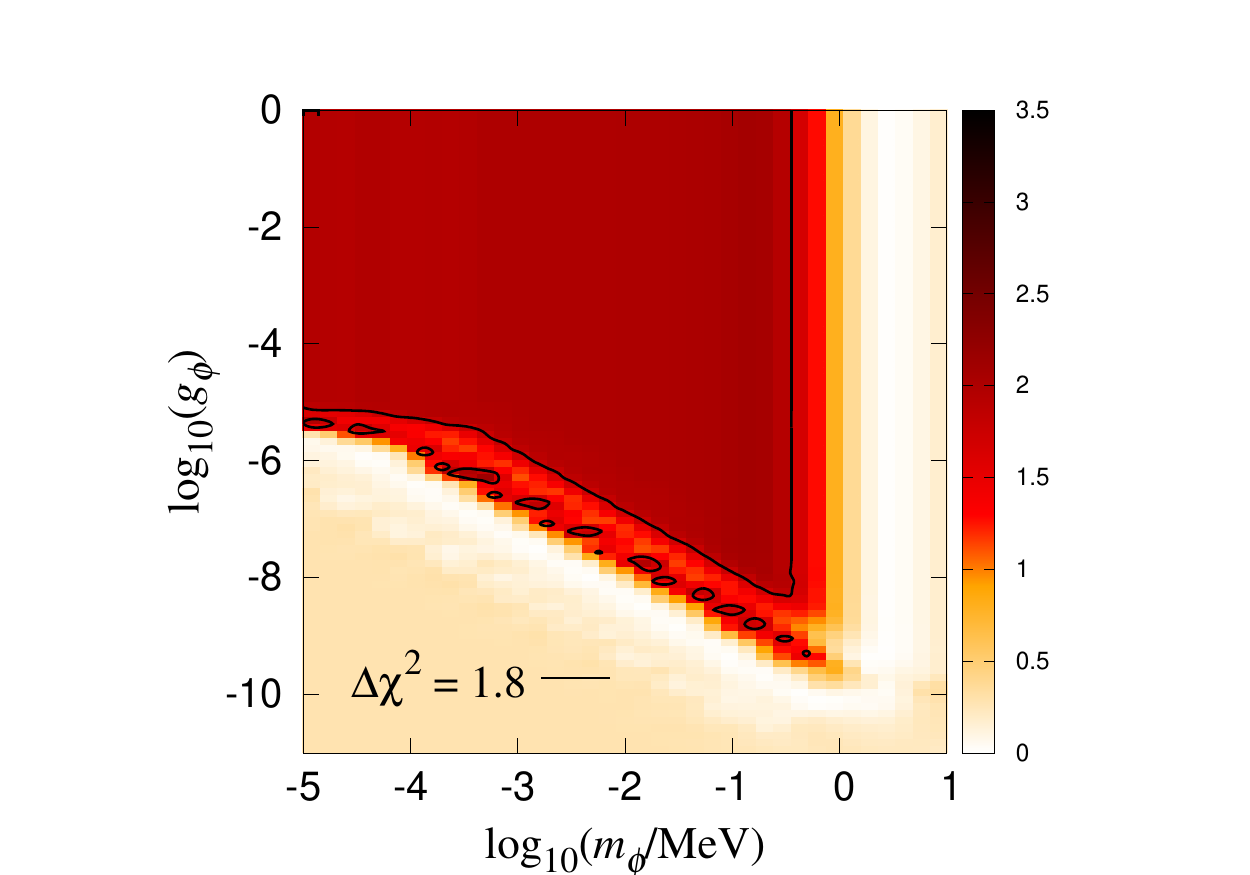}        }%
    \subfigure{%
    \hspace{-3cm}
    \includegraphics[width=0.65\textwidth]{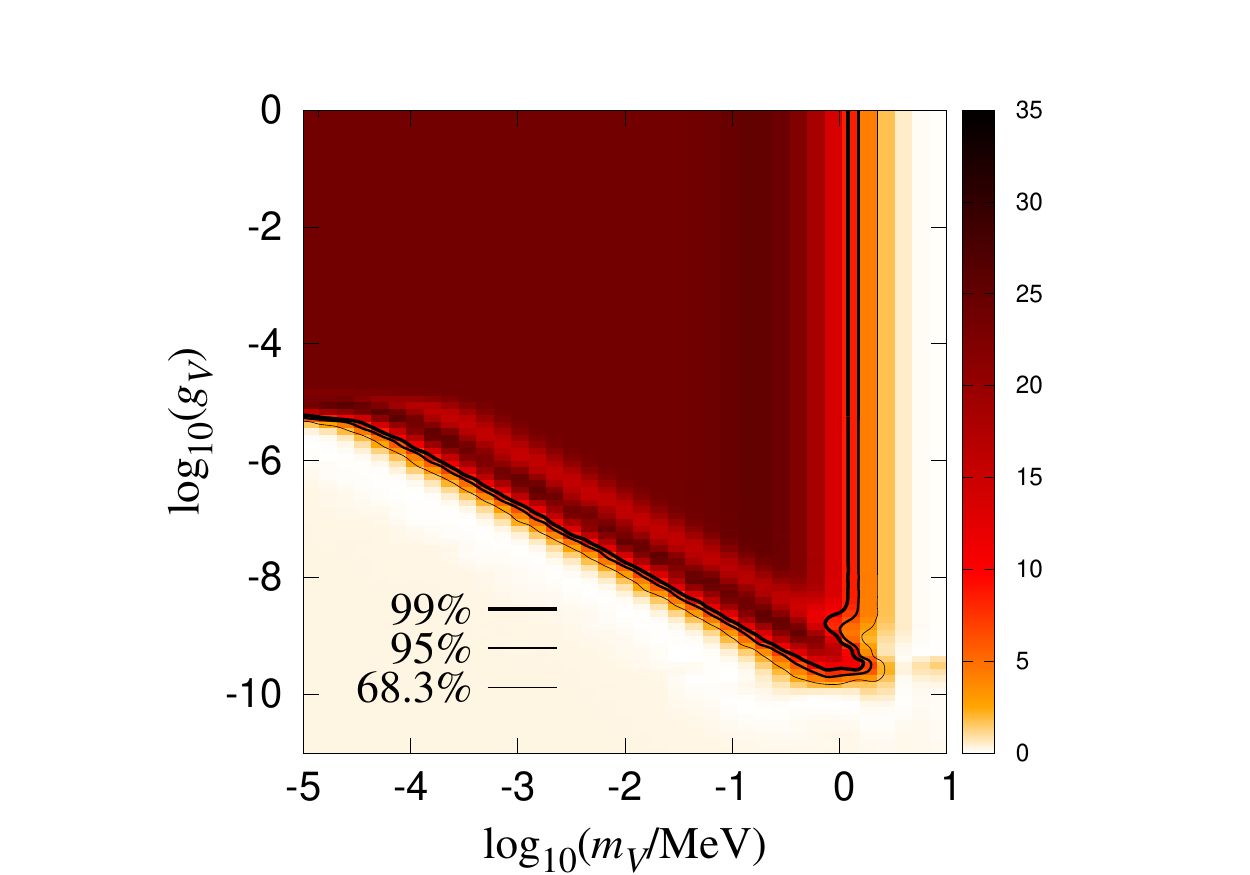}        }
    \end{center}
    \vspace{-0.8cm}
    \caption{The contour plot and density map of the $\chi^2$ function in the scalar-boson case (left panel) and those in the vector-boson case (right panel). Only the primordial mass fraction of helium $Y^{}_{\rm p} = 0.2449 \pm 0.0040$ and the primordial abundance of deuterium ${\rm D}/{\rm H}|^{}_{\rm p} = (2.53 \pm 0.04) \times 10^{-5}$ have been used in the statistical analysis.}
    \label{fig:bbn_constraints}
    \end{figure}

The ultimate goal of this work is to constrain the parameter space by the observations of the primordial abundances of deuterium and helium. For this purpose, we have to confront the theoretical predictions of these quantities with the observed values. With the help of our modified version of the AlterBBN code, we are able to calculate the light element abundances $Y^{}_i(\eta, g^{}_{\rm s}, m^{}_{\rm s})$ given the baryon-to-photon ratio $\eta$, the coupling constant $g^{}_{\rm s}$, and the mass $m^{}_{\rm s}$. Here the subscript ``s" denotes either $\phi$ for the scalar boson or $V$ for the vector boson. Apart from these input parameters, we need to take into account the theoretical errors on the nuclear reaction rates and the neutron lifetime. This is important, since the observational errors are currently comparable to the theoretical ones. To this end, we adopt the simple method of linear error propagation from Ref.~\cite{Fiorentini:1998fv} and define the $\chi^2$ function as
\begin{equation}	
\chi^2 = \sum_{i,j} (Y^{\rm th}_i - Y^{\rm ex}_i)[S^{}_{ij}]^{-1}(Y^{\rm th}_j - Y^{\rm ex}_j) \; ,
\label{eq:chi2}
\end{equation}
where $S^{}_{ij} \equiv \left(\sigma^2_{\rm th}\right)^{}_{ij} + \left(\sigma^2_{\rm ex}\right)^{}_{ij}$ is the covariance matrix of squared errors and the indices $i$ and $j$ refer to the light elements of our interest. We only consider the well-measured abundances of deuterium and helium, i.e., $i$ and $j$ run over ${\rm D}$ and ${^4}{\rm He}$. In Eq.~\eqref{eq:chi2}, $Y^{\rm th}_i$ and $Y^{\rm ex}_i$ are the theoretical predictions and the experimental values of the abundances, respectively. In our calculations, the theoretical errors $\left(\sigma^2_{\rm th}\right)^{}_{ij}$ are estimated by using the method of linear error propagation. For simplicity, we just consider the errors of the twelve most relevant nuclear reactions~\cite{Fiorentini:1998fv}. The theoretical errors on those nuclear reactions can be found in Refs.~\cite{Serpico:2004gx, Pisanti:2007hk, Arbey:2011nf}, where the astrophysical $S$ factors and their corresponding uncertainties are quantified as polynomial functions of temperature. One should be referred to Refs.~\cite{Marcucci:2015yla, Coc:2015bhi, Nollett:2011aa, Coc:2014oia, Dubovichenko:2017bmz, Adelberger:2010qa, Xu:2013fha, Pitrou:2018cgg, Cyburt:2008kw, Coc:2011az, Boyd:2010kj} for recent developments in this aspect. On the other hand, the experimental errors are assumed to be uncorrelated, namely, $\left(\sigma^2_{\rm ex}\right)^{}_{ij} = \delta^{}_{ij} \sigma^{i}_{\rm ex} \sigma^{j}_{\rm ex}$, where the standard deviation $\sigma^i_{\rm ex}$ is assumed for the corresponding individual observable. In Ref.~\cite{Aver:2015iza}, the newly added ${\rm He}~{\rm I}~\lambda 10830$ infrared emission lines have been incorporated together with the traditional visible lines from the metal-poor ionized hydrogen regions of compact blue galaxies to obtain $Y^{}_{\rm p} = 0.2449 \pm 0.0040$. The relative abundance of deuterium ${\rm D/H}|^{}_{\rm p} = (2.53 \pm 0.04) \times 10^{-5}$ is inferred from the observation of high-redshift interstellar clouds in Ref.~\cite{Cooke:2013cba}, in which all existing data have been systematically studied.

Then, we are ready to compute $\chi^2(\eta, g^{}_{\rm s}, m^{}_{\rm s})$ by scanning over the parameter space of $\eta$, $g^{}_{\rm s}$, and $m^{}_{\rm s}$. Since we are interested in the BBN constraints on $g^{}_{\rm s}$ and $m^{}_{\rm s}$, the values of $\chi^2(\eta, g^{}_{\rm s}, m^{}_{\rm s})$ are minimized with respect to $\eta$ in the range from $10^{-9}$ to $10^{-10}$. Since the SNI may also affect the formation of the CMB, the observational information on $\eta$ from the CMB data is not used for self-consistency. Since the abundance of deuterium is very sensitive to $\eta$ and the one of ${}^4{\rm He}$ is not, minimizing $\chi^2$ with respect to $\eta$ is equivalent to minimizing the deuterium deviation with respect to $\eta$. This is also why the deviations of the abundance of deuterium from the standard values are quite small in Fig.~\ref{fig:evolution} and Fig.~\ref{fig:evolution2} when the minimization is performed. In Fig.~\ref{fig:bbn_constraints}, the final results are presented, where the contours and the density map of the $\chi^2$ function have been plotted for the scalar-boson case (left panel) and the vector-boson case (right panel). In general, the excluded regions are similar to those in Fig.~\ref{fig:phi_contour}, which are basically described by simple arguments of $\Delta N^{}_{\rm eff}$. Nevertheless, some special features in Fig.~\ref{fig:bbn_constraints} can only be explained after a full calculation of light element abundances is accomplished.

In the left panel of Fig.~\ref{fig:bbn_constraints}, there is a gap between the small scattered circles and the boundary of a large connected region, corresponding to the contours of $\Delta\chi^2 = 1.8$. The reason is essentially the same as already mentioned in Case~III of the previous subsection. It is the freezing-in of $\phi$ before or after the neutrino decoupling that makes the key difference. In the former case, when $\phi$ is in thermal equilibrium before the neutrino decoupling, a large value of $g^{}_{\phi}$ is required. Thus, $T^{}_\gamma$ and $T^{}_\nu$ must be changed simultaneously such that their ratio is almost the same as that of the standard case. Hence, the light abundances are affected by the extra radiation, or equivalently, a large expansion rate. However, in the latter case, when $\phi$ freezes in after the neutrino decoupling, a smaller $g^{}_\phi$ is needed. Although the contribution to the extra radiation is small, the production of $\phi$'s from $\nu$'s will reduce $T^{}_\nu$, leading to an earlier decoupling of the weak interaction. Therefore, for a fixed $m^{}_\phi$, both a large coupling and a small one can give rise to the same $\chi^2$ function, successfully explaining the observed features. In the right panel of Fig.~\ref{fig:bbn_constraints}, the gap does not show up, but one can notice a dark region along the $99~\%$ contour in the density map, which has the same origin as the appearance of the smaller circles in the left panel.

With the above detailed calculations, we now make some remarks on the flavor dependence of the SNI and discuss its impact on the final observational constraints. If the mediator $\phi$ or $V$ is only coupled to muon and tau neutrinos (antineutrinos), the influence on BBN will be just the modification of the total energy density, which can entirely be represented by $N^{}_{\rm eff}$. However, if the mediator is coupled to the electron neutrinos (antineutrinos), the average energy or temperature of $\nu^{}_e$ ($\overline{\nu}^{}_e$) will be changed via the production of the mediator, which will affect the total energy density and directly modify the rates of weak interactions $\nu^{}_e + n \leftrightarrow p + e^-$ and $\overline{\nu}^{}_e + p \leftrightarrow n + e^+$ that are responsible for the $n/p$ ratio. As for the final constraints on the couplings and the mediator masses, the main difference between the electron and muon (or tau) neutrinos can be clearly observed by comparing the numerical results shown in Fig.~\ref{fig:bbn_constraints} and those in Fig.~\ref{fig:phi_contour}. In the former case, the dark regions along the indicated contours in Fig.~\ref{fig:bbn_constraints} signify the additional impact on the $n/p$ ratio in the case of the SNI for $\nu^{}_e$.

\section{Summary and Conclusions}
\label{sec:sc}
We have performed a detailed study of the SNI in the BBN era. In order to better understand the production and the evolution of the involved new scalar or vector boson, we have implemented the Boltzmann equations for the distribution functions of neutrinos and this new particle. Furthermore, these Boltzmann equations have been combined with the AlterBBN code to analyze the effects of the SNI in the evolution of light element abundances. The observed primordial abundances of deuterium and helium have been used to constrain the coupling and the mass of the new particle. Such a study is very helpful in exploring the intrinsic properties of massive neutrinos, including the origin of neutrino masses and the interactions among neutrinos, and in constraining other new-physics scenarios beyond the standard model.

In the scalar-boson case, it has been demonstrated that the BBN bound on $g^{}_\phi$ and $m^{}_\phi$ is very weak. However, in the vector-boson case, very stringent bounds on $g^{}_V$ can be obtained for a wide range of masses, i.e., $10^{-5}~{\rm MeV} \lesssim m^{}_V \lesssim 5~{\rm MeV}$. The most stringent bound $g^{}_V \lesssim 6\times 10^{-10}$ at $95~\%$ CL is achieved for $m^{}_V \simeq 1~{\rm MeV}$. The bound on $g^{}_V$ will be significantly relaxed for much smaller masses, namely, $g^{}_V \lesssim 8\times 10^{-6}$ for $m^{}_V \lesssim 10^{-4}~{\rm MeV}$ at the same CL. The BBN constraint is comparable to the supernova bound~\cite{Heurtier:2016otg}, but weaker than the bounds drawn from decays of kaon and weak gauge bosons in the low-mass region~\cite{Laha:2013xua}.

It is worthwhile to emphasize that the CMB data will be very useful in constraining the SNI. For instance, although the scalar boson $\phi$ can at most contribute an extra radiation of $\Delta N^{}_{\rm eff} = 0.57$, the CMB bound on the effective number of neutrino species is $N^{}_{\rm eff} = 3.14^{+0.44}_{-0.43}$ at $95~\%$ CL~\cite{Ade:2015xua}. In addition, when the temperature drops below $m^{}_\phi$, the Boltzmann suppression processes conserve the entropy, but not the comoving energy density, enhancing the extra radiation to $\Delta N^{}_{\rm eff} \approx 0.75$ in the CMB epoch. Similar arguments can be applied to the vector boson $V$ as well. However, it should be noted that the SNI in the CMB epoch may alter the power spectrum of anisotropy via a tight coupling~\cite{Hannestad:2004qu, Hannestad:2005ex, Bell:2005dr, Basboll:2008fx, Archidiacono:2013dua, Forastieri:2015paa, Cyr-Racine:2013jua, Oldengott:2014qra, Oldengott:2017fhy}. Fortunately, both a larger $N^{}_{\rm eff}$ and a stronger SNI enhance the power spectrum of anisotropy, leading to more restrictive constraints.

\section*{Acknowledgements}
The authors are indebted to Newton Nath for technical help in drawing the figures and Yasaman Farzan for useful discussions. The Feynman diagrams in Fig.~\ref{fig:feynman} have been generated by using the software Jaxodraw~\cite{Binosi:2003yf}. This work was in part supported by the National Natural Science Foundation of China under Grant No. 11775232, by the National Recruitment Program for Young Professionals and the CAS Center for Excellence in Particle Physics (CCEPP). T.O.~acknowledges support by the Swedish Research Council (Vetenskapsr\r{a}det) through contract No.~2017-03934 and the KTH Royal Institute of Technology for a sabbatical period at the University of Iceland.



\begin{thebibliography}{99}
\bibitem{Wagoner:1966pv}
  R.~V.~Wagoner, W.~A.~Fowler and F.~Hoyle,
  ``On the synthesis of elements at very high temperatures,''
  Astrophys.\ J.\  {\bf 148}, 3 (1967).

\bibitem{Boesgaard:1985km}
  A.~M.~Boesgaard and G.~Steigman,
  ``Big Bang Nucleosynthesis: Theories and Observations,''
  Ann.\ Rev.\ Astron.\ Astrophys.\  {\bf 23}, 319 (1985).

\bibitem{Schramm:1997vs}
  D.~N.~Schramm and M.~S.~Turner,
  ``Big-bang nucleosynthesis enters the precision era,''
  Rev.\ Mod.\ Phys.\  {\bf 70}, 303 (1998)
  [arXiv:astro-ph/9706069].

\bibitem{Olive:1999ij}
  K.~A.~Olive, G.~Steigman and T.~P.~Walker,
  ``Primordial nucleosynthesis: theory and observations,''
  Phys.\ Rep.\  {\bf 333}, 389 (2000)
  [arXiv:astro-ph/9905320].

\bibitem{Iocco:2008va}
  F.~Iocco, G.~Mangano, G.~Miele, O.~Pisanti and P.~D.~Serpico,
  ``Primordial nucleosynthesis: From precision cosmology to fundamental physics,''
  Phys.\ Rep.\  {\bf 472}, 1 (2009)
  [arXiv:0809.0631].

\bibitem{Cyburt:2015mya}
  R.~H.~Cyburt, B.~D.~Fields, K.~A.~Olive and T.~H.~Yeh,
  ``Big bang nucleosynthesis: Present status,''
  Rev.\ Mod.\ Phys.\  {\bf 88}, 015004 (2016)
  [arXiv:1505.01076].

\bibitem{Sarkar:1995dd}
  S.~Sarkar,
  ``Big bang nucleosynthesis and physics beyond the standard model,''
  Rep.\ Prog.\ Phys.\  {\bf 59}, 1493 (1996)
  [arXiv:hep-ph/9602260].

\bibitem{Pospelov:2010hj}
  M.~Pospelov and J.~Pradler,
  ``Big Bang Nucleosynthesis as a Probe of New Physics,''
  Ann.\ Rev.\ Nucl.\ Part.\ Sci.\  {\bf 60}, 539 (2010)
  [arXiv:1011.1054].

\bibitem{Fields:2011zzb}
  B.~D.~Fields,
  ``The Primordial Lithium Problem,''
  Ann.\ Rev.\ Nucl.\ Part.\ Sci.\  {\bf 61}, 47 (2011)
  [arXiv:1203.3551].

\bibitem{Ade:2015xua}
  P.~A.~R.~Ade {\it et al.} [Planck Collaboration],
  ``Planck 2015 results. XIII. Cosmological parameters,''
  Astron.\ Astrophys.\  {\bf 594}, A13 (2016)
  [arXiv:1502.01589].

\bibitem{Aver:2015iza}
  E.~Aver, K.~A.~Olive and E.~D.~Skillman,
  ``The effects of He I $\lambda$10830 on helium abundance determinations,''
  J.\ Cosmol.\ Astropart.\ Phys. {\bf 07}, 011 (2015)
  [arXiv:1503.08146].

\bibitem{Cooke:2013cba}
  R.~Cooke, M.~Pettini, R.~A.~Jorgenson, M.~T.~Murphy and C.~C.~Steidel,
  ``Precision measures of the primordial abundance of deuterium,''
  Astrophys.\ J.\  {\bf 781}, 31 (2014)
  [arXiv:1308.3240].

\bibitem{Bania:2002yj}
  T.~M.~Bania, R.~T.~Rood and D.~S.~Balser,
  ``The cosmological density of baryons from observations of ${}^3{\rm He}^+$ in the Milky Way,''
  Nature {\bf 415}, 54 (2002).

\bibitem{Sbordone:2010zi}
  L.~Sbordone {\it et al.},
  ``The metal-poor end of the Spite plateau. I. Stellar parameters, metallicities, and lithium abundances,''
  Astron.\ Astrophys.\  {\bf 522}, A26 (2010)
  [arXiv:1003.4510].

\bibitem{Fields:2014uja}
  B.~D.~Fields, P.~Molaro and S.~Sarkar,
  ``Big-Bang Nucleosynthesis,''
  Chin.\ Phys.\ C {\bf 38}, 339 (2014)
  [arXiv:1412.1408].

\bibitem{Patrignani:2016xqp}
  C.~Patrignani {\it et al.} [Particle Data Group],
  ``Review of Particle Physics,''
  Chin.\ Phys.\ C {\bf 40}, 100001 (2016).

\bibitem{Singh:2017jmz}
  V.~Singh, J.~Lahiri, D.~Bhowmick and D.~N.~Basu,
  ``Primordial lithium abundance problem of BBN and baryonic density in the universe,''
  arXiv:1708.05567.

\bibitem{Dolgov:2002wy}
  A.~D.~Dolgov,
  ``Neutrinos in cosmology,''
  Phys.\ Rep.\  {\bf 370}, 333 (2002)
  [arXiv:hep-ph/0202122].

\bibitem{Hannestad:2006zg}
  S.~Hannestad,
  ``Primordial Neutrinos,''
  Ann.\ Rev.\ Nucl.\ Part.\ Sci.\  {\bf 56}, 137 (2006)
  [arXiv:hep-ph/0602058].

\bibitem{Lesgourgues:2006nd}
  J.~Lesgourgues and S.~Pastor,
  ``Massive neutrinos and cosmology,''
  Phys.\ Rep.\  {\bf 429}, 307 (2006)
  [arXiv:astro-ph/0603494].

\bibitem{Wong:2011ip}
  Y.~Y.~Y.~Wong,
  ``Neutrino Mass in Cosmology: Status and Prospects,''
  Ann.\ Rev.\ Nucl.\ Part.\ Sci.\  {\bf 61}, 69 (2011)
  [arXiv:1111.1436].

\bibitem{Raffelt:1996wa}
  G.~G.~Raffelt,
  ``Stars as laboratories for fundamental physics: The astrophysics of neutrinos, axions, and other weakly interacting particles,''
  Chicago, USA: Univ. Pr. (1996) 664 p.

\bibitem{Chikashige:1980ui}
  Y.~Chikashige, R.~N.~Mohapatra and R.~D.~Peccei,
  ``Are there real goldstone bosons associated with broken lepton number?,''
  Phys.\ Lett.\ B {\bf 98}, 265 (1981).

\bibitem{Gelmini:1980re}
  G.~B.~Gelmini and M.~Roncadelli,
  ``Left-handed neutrino mass scale and spontaneously broken lepton number,''
  Phys.\ Lett.\ B {\bf 99}, 411 (1981).

\bibitem{Choi:1991aa}
  K.~Choi and A.~Santamaria,
  ``17~keV neutrino in a singlet-triplet majoron model,''
  Phys.\ Lett.\ B {\bf 267}, 504 (1991).

\bibitem{Acker:1992eh}
  A.~Acker, A.~Joshipura and S.~Pakvasa,
  ``A neutrino decay model, solar antineutrinos and atmospheric neutrinos,''
  Phys.\ Lett.\ B {\bf 285}, 371 (1992).

\bibitem{Georgi:1981pg}
  H.~M.~Georgi, S.~L.~Glashow and S.~Nussinov,
  ``Unconventional model of neutrino masses,''
  Nucl.\ Phys.\ B {\bf 193}, 297 (1981).

\bibitem{Schechter:1981cv}
  J.~Schechter and J.~W.~F.~Valle,
  ``Neutrino decay and spontaneous violation of lepton number,''
  Phys.\ Rev.\ D {\bf 25}, 774 (1982).

\bibitem{Aarssen:2012fx}
  L.~G.~van den Aarssen, T.~Bringmann and C.~Pfrommer,
  ``Is Dark Matter with Long-Range Interactions a Solution to All Small-Scale Problems of $\Lambda$ Cold Dark Matter Cosmology?,''
  Phys.\ Rev.\ Lett.\  {\bf 109}, 231301 (2012)
  [arXiv:1205.5809].


\bibitem{Hannestad:2004qu}
  S.~Hannestad,
  ``Structure formation with strongly interacting neutrinos---implications for the cosmological neutrino mass bound,''
  J.\ Cosmol.\ Astropart.\ Phys. {\bf 02}, 011 (2005)
  [arXiv:astro-ph/0411475].

\bibitem{Hannestad:2005ex}
  S.~Hannestad and G.~Raffelt,
  ``Constraining invisible neutrino decays with the cosmic microwave background,''
  Phys.\ Rev.\ D {\bf 72}, 103514 (2005)
  [arXiv:hep-ph/0509278].

\bibitem{Bell:2005dr}
  N.~F.~Bell, E.~Pierpaoli and K.~Sigurdson,
  ``Cosmological signatures of interacting neutrinos,''
  Phys.\ Rev.\ D {\bf 73}, 063523 (2006)
  [arXiv:astro-ph/0511410].

\bibitem{Basboll:2008fx}
  A.~Basb{\o}ll, O.~E.~Bjaelde, S.~Hannestad and G.~G.~Raffelt,
  ``Are cosmological neutrinos free-streaming?,''
  Phys.\ Rev.\ D {\bf 79}, 043512 (2009)
  [arXiv:0806.1735].

\bibitem{Archidiacono:2013dua}
  M.~Archidiacono and S.~Hannestad,
  ``Updated constraints on non-standard neutrino interactions from Planck,''
  J.\ Cosmol.\ Astropart.\ Phys. {\bf 07}, 046 (2014)
  [arXiv:1311.3873].

\bibitem{Forastieri:2015paa}
  F.~Forastieri, M.~Lattanzi and P.~Natoli,
  ``Constraints on secret neutrino interactions after Planck,''
  J.\ Cosmol.\ Astropart.\ Phys. {\bf 07}, 014 (2015)
  [arXiv:1504.04999].

\bibitem{Cyr-Racine:2013jua}
  F.~Y.~Cyr-Racine and K.~Sigurdson,
  ``Limits on neutrino-neutrino scattering in the early Universe,''
  Phys.\ Rev.\ D {\bf 90}, 123533 (2014)
  [arXiv:1306.1536].

\bibitem{Oldengott:2014qra}
  I.~M.~Oldengott, C.~Rampf and Y.~Y.~Y.~Wong,
  ``Boltzmann hierarchy for interacting neutrinos I: formalism,''
  J.\ Cosmol.\ Astropart.\ Phys. {\bf 04}, 016 (2015)
  [arXiv:1409.1577].

\bibitem{Forastieri:2017oma}
  F.~Forastieri, M.~Lattanzi, G.~Mangano, A.~Mirizzi, P.~Natoli and N.~Saviano,
  ``Cosmic microwave background constraints on secret interactions among sterile neutrinos,''
  J.\ Cosmol.\ Astropart.\ Phys. {\bf 07}, 038 (2017)
  [arXiv:1704.00626].

\bibitem{Lancaster:2017ksf}
  L.~Lancaster, F.~Y.~Cyr-Racine, L.~Knox and Z.~Pan,
  ``A tale of two modes: neutrino free-streaming in the early universe,''
  J.\ Cosmol.\ Astropart.\ Phys. {\bf 07}, 033 (2017)
  [arXiv:1704.06657].

\bibitem{Oldengott:2017fhy}
  I.~M.~Oldengott, T.~Tram, C.~Rampf and Y.~Y.~Y.~Wong,
  ``Interacting neutrinos in cosmology: exact description and constraints,''
  J.\ Cosmol.\ Astropart.\ Phys. {\bf 11}, 027 (2017)
  [arXiv:1706.02123].

\bibitem{Kolb:1987qy}
  E.~W.~Kolb and M.~S.~Turner,
  ``Supernova 1987A and the secret interactions of neutrinos,''
  Phys.\ Rev.\ D {\bf 36}, 2895 (1987).

\bibitem{Konoplich:1988mj}
  R.~V.~Konoplich and M.~Y.~Khlopov,
  ``Constraints on triplet Majoron model due to observations of neutrinos from stellar collapse,''
  Sov.\ J.\ Nucl.\ Phys.\  {\bf 47}, 565 (1988)
  [Yad.\ Fiz.\  {\bf 47}, 891 (1988)].

\bibitem{Farzan:2002wx}
  Y.~Farzan,
  ``Bounds on the coupling of the Majoron to light neutrinos from supernova cooling,''
  Phys.\ Rev.\ D {\bf 67}, 073015 (2003)
  [arXiv:hep-ph/0211375].

\bibitem{Zhou:2011rc}
  S.~Zhou,
  ``Comment on `Astrophysical consequences of a neutrinophilic two-Higgs-doublet model',''
  Phys.\ Rev.\ D {\bf 84}, 038701 (2011)
  [arXiv:1106.3880].


\bibitem{Heurtier:2016otg}
  L.~Heurtier and Y.~Zhang,
  ``Supernova constraints on massive (pseudo)scalar coupling to neutrinos,''
  J.\ Cosmol.\ Astropart.\ Phys. {\bf 02}, 042 (2017)
  [arXiv:1609.05882].

\bibitem{Ng:2014pca}
  K.~C.~Y.~Ng and J.~F.~Beacom,
  ``Cosmic neutrino cascades from secret neutrino interactions,''
  Phys.\ Rev.\ D {\bf 90}, 065035 (2014)
  Erratum: [Phys.\ Rev.\ D {\bf 90}, 089904 (2014)]
  [arXiv:1404.2288].

\bibitem{Ioka:2014kca}
  K.~Ioka and K.~Murase,
  ``IceCube PeV--EeV neutrinos and secret interactions of neutrinos,''
  Progr.\ Theor.\ Exp.\ Phys. {\bf 2014}, 061E01 (2014)
  [arXiv:1404.2279].

\bibitem{Gando:2012pj}
  A.~Gando {\it et al.} [KamLAND-Zen Collaboration],
  ``Limits on Majoron-emitting double-$\beta$ decays of ${}^{136}{\rm Xe}$ in the KamLAND-Zen experiment,''
  Phys.\ Rev.\ C {\bf 86}, 021601 (2012)
  [arXiv:1205.6372].

\bibitem{Albert:2014fya}
  J.~B.~Albert {\it et al.} [EXO-200 Collaboration],
  ``Search for Majoron-emitting modes of double-beta decay of $^{136}{\rm Xe}$ with EXO-200,''
  Phys.\ Rev.\ D {\bf 90}, 092004 (2014)
  [arXiv:1409.6829].

\bibitem{Belotsky:2001fb}
  K.~M.~Belotsky, A.~L.~Sudarikov and M.~Y.~Khlopov,
  ``Constraint on anomalous 4nu interaction,''
  Phys.\ Atom.\ Nucl.\  {\bf 64}, 1637 (2001)
  [Yad.\ Fiz.\  {\bf 64}, 1718 (2001)].

\bibitem{Laha:2013xua}
  R.~Laha, B.~Dasgupta and J.~F.~Beacom,
  ``Constraints on new neutrino interactions via light Abelian vector bosons,''
  Phys.\ Rev.\ D {\bf 89}, 093025 (2014)
  [arXiv:1304.3460].

\bibitem{Ahlgren:2013wba}
  B.~Ahlgren, T.~Ohlsson and S.~Zhou,
  ``Comment on `Is Dark Matter with Long-Range Interactions a Solution to All Small-Scale Problems of $\Lambda$ Cold Dark Matter Cosmology?',''
  Phys.\ Rev.\ Lett.\  {\bf 111}, 199001 (2013)
  [arXiv:1309.0991].

\bibitem{Mangano:2005cc}
  G.~Mangano, G.~Miele, S.~Pastor, T.~Pinto, O.~Pisanti and P.~D.~Serpico,
  ``Relic neutrino decoupling including flavour oscillations,''
  Nucl.\ Phys.\ B {\bf 729}, 221 (2005)
  [arXiv:hep-ph/0506164].

\bibitem{Birrell:2014uka}
  J.~Birrell, C.~T.~Yang and J.~Rafelski,
  ``Relic neutrino freeze-out: Dependence on natural constants,''
  Nucl.\ Phys.\ B {\bf 890}, 481 (2014)
  [arXiv:1406.1759].

\bibitem{Grohs:2015tfy}
  E.~Grohs, G.~M.~Fuller, C.~T.~Kishimoto, M.~W.~Paris and A.~Vlasenko,
  ``Neutrino energy transport in weak decoupling and big bang nucleosynthesis,''
  Phys.\ Rev.\ D {\bf 93}, 083522 (2016)
  [arXiv:1512.02205].

\bibitem{deSalas:2016ztq}
  P.~F.~de Salas and S.~Pastor,
  ``Relic neutrino decoupling with flavour oscillations revisited,''
  J.\ Cosmol.\ Astropart.\ Phys. {\bf 07}, 051 (2016)
  [arXiv:1606.06986].

\bibitem{Mangano:2011ar}
  G.~Mangano and P.~D.~Serpico,
  ``A robust upper limit on $N_{\rm eff}$ from BBN, circa 2011,''
  Phys.\ Lett.\ B {\bf 701}, 296 (2011)
  [arXiv:1103.1261].

\bibitem{Kolb:1990vq}
  E.~W.~Kolb and M.~S.~Turner,
  ``The Early Universe,''
  Front.\ Phys.\  {\bf 69}, 1 (1990).

\bibitem{Dodelson:2003ft}
  S.~Dodelson,
  ``Modern Cosmology,''
  Amsterdam, Netherlands: Academic Pr. (2003) 440 p.

\bibitem{Weinberg:2008zzc}
  S.~Weinberg,
  ``Cosmology,''
  Oxford, UK: Oxford Univ. Pr. (2008) 593 p.

\bibitem{Bernstein:1988bw}
  J.~Bernstein,
  ``Kinetic Theory in The Expanding Universe,''
  Cambridge, UK: Cambridge Univ. Pr. (1988) 149p.

\bibitem{Hannestad:1995rs}
  S.~Hannestad and J.~Madsen,
  Phys.\ Rev.\ D {\bf 52}, 1764 (1995)
  [astro-ph/9506015].

\bibitem{Dolgov:1997mb}
  A.~D.~Dolgov, S.~H.~Hansen and D.~V.~Semikoz,
  Nucl.\ Phys.\ B {\bf 503}, 426 (1997)
  [hep-ph/9703315].

\bibitem{Esposito:2000hi}
  S.~Esposito, G.~Miele, S.~Pastor, M.~Peloso and O.~Pisanti,
  Nucl.\ Phys.\ B {\bf 590}, 539 (2000)
  [astro-ph/0005573].

\bibitem{Arbey:2011nf}
  A.~Arbey,
  ``AlterBBN: A program for calculating the BBN abundances of the elements in alternative cosmologies,''
  Comput.\ Phys.\ Commun.\  {\bf 183}, 1822 (2012)
  [arXiv:1106.1363].

\bibitem{Wagoner:1969}
  R.~V.~Wagoner,
  ``Synthesis of the elements within objects exploding from very high temperatures,''
  Astrophys.\ J.\ Suppl.\  {\bf 18}, 247 (1969).

\bibitem{Kawano:1992ua}
  L.~Kawano,
  ``Let's go: Early universe II. Primordial Nucleosynthesis: The Computer Way,''
  FERMILAB-PUB-92-004-A.

\bibitem{Pisanti:2007hk}
  O.~Pisanti, A.~Cirillo, S.~Esposito, F.~Iocco, G.~Mangano, G.~Miele and P.~D.~Serpico,
  ``PArthENoPE: Public algorithm evaluating the nucleosynthesis of primordial elements,''
  Comput.\ Phys.\ Commun.\  {\bf 178}, 956 (2008)
  [arXiv:0705.0290].

\bibitem{Consiglio:2017pot}
  R.~Consiglio, P.~F.~de Salas, G.~Mangano, G.~Miele, S.~Pastor and O.~Pisanti,
  ``PArthENoPE reloaded,''
  arXiv:1712.04378.

\bibitem{Fiorentini:1998fv}
  G.~Fiorentini, E.~Lisi, S.~Sarkar and F.~L.~Villante,
  ``Quantifying uncertainties in primordial nucleosynthesis without Monte Carlo simulations,''
  Phys.\ Rev.\ D {\bf 58}, 063506 (1998)
  [arXiv:astro-ph/9803177].

\bibitem{Serpico:2004gx}
  P.~D.~Serpico, S.~Esposito, F.~Iocco, G.~Mangano, G.~Miele and O.~Pisanti,
  ``Nuclear reaction network for primordial nucleosynthesis: A Detailed analysis of rates, uncertainties and light nuclei yields,''
  JCAP {\bf 0412}, 010 (2004)
   [astro-ph/0408076].




\bibitem{Marcucci:2015yla}
  L.~E.~Marcucci, G.~Mangano, A.~Kievsky and M.~Viviani,
  ``Implication of the proton-deuteron radiative capture for Big Bang Nucleosynthesis,''
  Phys.\ Rev.\ Lett.\  {\bf 116}, 102501 (2016)
  Erratum: [Phys.\ Rev.\ Lett.\  {\bf 117}, 049901 (2016)]
  [arXiv:1510.07877].

\bibitem{Coc:2015bhi}
  A.~Coc, P.~Petitjean, J.~P.~Uzan, E.~Vangioni, P.~Descouvemont, C.~Iliadis and R.~Longland,
  ``New reaction rates for improved primordial D/H calculation and the cosmic evolution of deuterium,''
  Phys.\ Rev.\ D {\bf 92}, 123526 (2015)
  [arXiv:1511.03843].

\bibitem{Nollett:2011aa}
  K.~M.~Nollett and G.~P.~Holder,
  ``An analysis of constraints on relativistic species from primordial nucleosynthesis and the cosmic microwave background,''
  arXiv:1112.2683.

\bibitem{Coc:2014oia}
  A.~Coc, J.~P.~Uzan and E.~Vangioni,
  ``Standard big bang nucleosynthesis and primordial CNO Abundances after Planck,''
  JCAP {\bf 1410}, 050 (2014)
  [arXiv:1403.6694].

\bibitem{Dubovichenko:2017bmz}
  S.~B.~Dubovichenko, A.~V.~Dzhazairov-Kakhramanov and N.~A.~Burkova,
  ``New Results for Astrophysical S-Factors of Radiative $^{3}$He$^{4}$He, $^{3}$H$^{4}$He, and $^{2}$H$^{4}$He Capture,''
  Russ.\ Phys.\ J.\  {\bf 60}, 1143 (2017)
  [arXiv:1706.05245].

\bibitem{Adelberger:2010qa}
  E.~G.~Adelberger {\it et al.},
  ``Solar fusion cross sections II: the pp chain and CNO cycles,''
  Rev.\ Mod.\ Phys.\  {\bf 83}, 195 (2011)
  [arXiv:1004.2318].

\bibitem{Xu:2013fha}
  Y.~Xu, K.~Takahashi, S.~Goriely, M.~Arnould, M.~Ohta and H.~Utsunomiya,
  ``NACRE II: an update of the NACRE compilation of charged-particle-induced thermonuclear reaction rates for nuclei with mass number $A < 16$,''
  Nucl.\ Phys.\ A {\bf 918}, 61 (2013)
  [arXiv:1310.7099].

\bibitem{Pitrou:2018cgg}
  C.~Pitrou, A.~Coc, J.~P.~Uzan and E.~Vangioni,
  ``Precision big bang nucleosynthesis with improved Helium-4 predictions,''
  arXiv:1801.08023.

\bibitem{Cyburt:2008kw}
  R.~H.~Cyburt, B.~D.~Fields and K.~A.~Olive,
  ``An Update on the big bang nucleosynthesis prediction for Li-7: The problem worsens,''
  JCAP {\bf 0811}, 012 (2008)
  [arXiv:0808.2818].

\bibitem{Coc:2011az}
  A.~Coc, S.~Goriely, Y.~Xu, M.~Saimpert and E.~Vangioni,
  ``Standard Big-Bang Nucleosynthesis up to CNO with an improved extended nuclear network,''
  Astrophys.\ J.\  {\bf 744}, 158 (2012)
  [arXiv:1107.1117].

\bibitem{Boyd:2010kj}
  R.~N.~Boyd, C.~R.~Brune, G.~M.~Fuller and C.~J.~Smith,
  ``New Nuclear Physics for Big Bang Nucleosynthesis,''
  Phys.\ Rev.\ D {\bf 82}, 105005 (2010)
  [arXiv:1008.0848].





\bibitem{Binosi:2003yf}
  D.~Binosi and L.~Theu{\ss}l,
  ``JaxoDraw: A graphical user interface for drawing Feynman diagrams,''
  Comput.\ Phys.\ Commun.\  {\bf 161}, 76 (2004)
  [arXiv:hep-ph/0309015].
\end{thebibliography}
\end{document}